\documentclass[12pt]{iopart}
\usepackage[utf8]{inputenc}
\usepackage{iopams}
\usepackage[colorlinks=true,hyperfootnotes=true,breaklinks=true,citecolor=red,urlcolor=blue,linkcolor=blue]{hyperref}
\usepackage{enumerate,graphicx,adjustbox,xcolor,tabularx}

\usepackage{cite} 
\usepackage{soul} 

\newtheorem{protocol}{Protocol}

\newcommand{\eqref}[1]{(\ref{#1})}
\newcommand{\ket}[1]{\left\vert {#1} \right\rangle}
\newcommand{\bra}[1]{\left\langle {#1} \right\vert}

\begin{document}

\title[Cyclic solid-state quantum battery]{Cyclic solid-state quantum battery: Thermodynamic characterization and quantum hardware simulation} 

\author{Luca Razzoli$^{1,2}$, Giulia Gemme$^{3,4}$, Ilia Khomchenko $^{1,2,5}$, Maura Sassetti$^{3,4}$, Henni Ouerdane$^{5}$, Dario Ferraro$^{3,4}$ and Giuliano Benenti$^{1,2}$}
    
\address{$^{1}$ Center for Nonlinear and Complex Systems, Dipartimento di Scienza e Alta Tecnologia, Universit\`a degli Studi dell'Insubria, Via Valleggio 11, 22100 Como, Italy}
\address{$^{2}$ Istituto Nazionale di Fisica Nucleare, Sezione di Milano, Via Celoria 16, 20133 Milano, Italy}
\address{$^{3}$ Dipartimento di Fisica, Universit\`a di Genova, Via Dodecaneso 33, 16146 Genova, Italy}
\address{$^{4}$ CNR-SPIN,  Via  Dodecaneso  33,  16146  Genova, Italy}
\address{$^{5}$ Digital Engineering Center, Skolkovo Institute of Science and Technology, 30 Bolshoi Boulevard, bld. 1, Moscow 121205, Russia}
\ead{\mailto{giulia.gemme@edu.unige.it}}

\begin{abstract}
We introduce a cyclic quantum battery model, based on an interacting bipartite system, weakly coupled to a thermal bath. The working cycle of the battery consists of four strokes: system thermalization, disconnection of subsystems, ergotropy extraction, and reconnection. The thermal bath acts as a charger in the thermalization stroke, while ergotropy extraction is possible because the ensuing thermal state is no longer passive after the disconnection stroke. Focusing on the case of two interacting qubits, we show that phase coherence, in the presence of non-trivial correlations between the qubits, can be exploited to reach working regimes with efficiency higher than $50\%$ while providing finite ergotropy. Our protocol is illustrated through a simple and feasible circuit model of a cyclic superconducting quantum battery. Furthermore, we simulate the considered cycle on superconducting IBM quantum machines. The good agreement between the theoretical and simulated results strongly suggests that our scheme for cyclic quantum batteries can be successfully realized in superconducting quantum hardware.
\end{abstract}

\maketitle

\section{Introduction}
Phase coherence and entanglement can be harnessed to provide powerful resources to boost the performance of quantum devices in comparison with their classical counterparts. Indeed, devices that rely on these phenomena can achieve a substantial advantage over classical machines in a wide range of technological application including secure data exchange~\cite{Gisin_2002, Portmann_2022}, computation~\cite{Benenti_book,Preskill18,Arute2019,Pan2020,Bluvstein2024},
and simulation of the behavior of complex systems under controlled conditions~\cite{Georgescu14, Daley2022}.

Turning specifically to power generation, the major challenges of nowadays technology concern the efficient conversion, storage and transport of energy. In this direction, batteries are clearly a promising solution~\cite{Hannan_2021}. However, despite undeniable advantages such as, e.g., high energy density, they suffer from intrinsic limitations like leakage and deteriorating state of health. For this reason, to overcome the current limitations new solutions still based on the principles of electrochemistry are actively explored, but another approach considers quantum mechanics as a new paradigm for energy harvesting, storage, and transfer. Interestingly, it has gained momentum in the recent years, albeit from a very fundamental and speculative standpoint. Studies retracing what was done for computation, with due distinctions, in the energetic case, paved the way a decade ago to the new and rapidly developing field of quantum batteries (QBs) -- see~\cite{Alicki_2013, Hovhannisyan_2013, Salimi2020,Binder_2015,Campaioli2017,Mitchison2021,Ju-Yeon2022,Quach_2023,Morrone2023qst,Fischer2024,Campaioli23} and references therein.

A first step towards possible experimental implementation of QBs was done by Ferraro \emph{et al.}~\cite{Ferraro_2018}, who proposed a scheme for a solid-state QB composed of a collection of two-level systems (TLSs) embedded in a resonant cavity, playing the role of quantum charger. This model has sparked interest in feasible models of QBs. Since then, various proposals of implementations on different platforms have appeared, including one-dimensional spin-chains~\cite{Le_2018, Dou22, Yao22, Crescente22, Grazi24}, strongly interacting fermionic systems~\cite{Rossini_2020, Rosa20, DeFilippis23}, and micromasers~\cite{Seah21, Shaghaghi22,Salvia23, shaghaghi23, Rodriguez23}. To date, there are very few actual realizations of QB prototypes; these are based on an organic microcavity~\cite{Quach_2022}, a nuclear spin-systems~\cite{Joshi_2022}, and a superconducting qutrit~\cite{Hu_2022, Gemme22, Gemme24}. 

The theoretical and experimental works cited above mainly considered the energy stored in the QB and the associated power, namely the energy divided by the charging time~\cite{Andolina18}. Despite their relevance for the advancement of the subject, they lack a more complete thermodynamic description of the operating regimes of these devices. To make progress, an interesting strategy consists in considering the operation of the QB as a thermodynamic cycle~\cite{Barra_njp, Hovhannisyan_prr}. There, the QB coupled to its charger is first typically initialized in a thermal state. Then the QB is disconnected from the charger (involving some energy cost). During the next step, the maximum amount of energy that can be extracted (known as ergotropy) is taken out from the QB through unitary operations. Note that energy extraction in this configuration is guaranteed by the fact that a thermalized state for the whole system (QB+charger) is not, in general, a passive thermal state for the QB only. In the final stage of the cycle, the QB and the charger are connected again (involving an additional energy cost) and the whole system is finally reinitialized, ready to perform a new cycle. Within this picture, the efficiency of the QB for each cycle can be defined as the ergotropy divided by the energy spent to carry out the connection/disconnection operations. In the original proposals~\cite{Barra_njp, Hovhannisyan_prr} the charger was assumed as a large system playing also the role of thermal bath (reservoir). This can represent an issue in view of possible implementations of the protocols, since it is not always possible in practice to connect/disconnect a quantum system from its environment at will. 
{In this regard, it has been recently shown that locally extracting work from a system in the presence of Hamiltonian couplings, e.g., with an environment, can deliver higher net work with respect to analogous protocols in which the interaction is switched off  \cite{Salvia23b,DiBello24,Castellano24,Castellano24b}.}

Our work proposes a 
{new} thermodynamic cycle in which an environment is constantly weakly coupled with the QB, but with the assumption that the operations are carried out on timescales shorter than those of thermalization as well as relaxation and dephasing of the QB. This ensures that the only role played by the bath is to reinitialize the system to the same thermal state at the end of each cycle. Moreover, as QB we consider a bipartite quantum system (specifically two coupled TLSs). In this setup it is possible to extract energy by means of local unitary operations on one or both TLSs separately or globally acting on them. 
We will show that, even if the ergotropy is larger in the latter case, the overall efficiency of the cycle can be higher acting with local unitary operations on the two TLSs, provided the same ergotropy. 
{Furthermore, we will show that there are parameter regions (i.e., low temperature and large enough TLS-TLS coupling strength) where the local energy extraction can outperform the global one, in terms of a single trade-off parameter encompassing both the amount of extracted work and the efficiency of the thermodynamic cycle.} 
This enhancement takes advantage of phase coherence in the presence of non-trivial correlations between the two TLSs. Since local unitary operations are typically easier to implement than global ones, this theoretical result opens interesting perspectives for the realization of cyclic QBs. 
{Note that here and in the following we use the adjective ``local'' to refer to the extractable work from non-interacting subsystems (condition achieved here after quenching off their interaction) as opposed to the local ergotropy intended as the extractable work from a system alone when interacting with an environment \cite{Salvia23b}.}

{Our study not only offers new theoretical insights into the thermodynamics of a cyclic quantum battery, but also presents a solid-state experimental proposal and a proof-of-principle simulation by means of real quantum hardware. Indeed, on} the one hand, we present a feasible scheme of a superconducting QB that can implement the proposed cycle, and we also provide typical experimental values for the circuit elements. Here, the two TLSs composing the QB are charge qubits coupled by means of an additional superconducting circuit \cite{Wang_prb} which allows for externally tuning the interaction. The use of superconducting circuits represents a major advantage, as they are nowadays a well-established solid-state platform~\cite{Krantz19,Kjaergaard2020} which has already proved its versatility in various contexts, with applications ranging from quantum information processing and computing~\cite{Wendin_2017} to sensing~\cite{Granata2016,Khomchenko2022,Vettoliere2023}, and in the context of hybrid quantum circuits for quantum technologies \cite{Blais2021}. 
On the other hand, we simulate the considered cycle on superconducting IBM quantum machines. Here, the most critical point is the realization of the initial thermal state for the system, since only unitary operations and measurements are available to end users on quantum computers.
This problem has been overcome by using the thermofield-double-state technique, i.e., by considering proper unitary operations acting on an enlarged Hilbert space and tracing then out the auxiliary degrees of freedom~\cite{Wu_prl2019,Zhu_pnas2020,Sagastizabal_npj2021,Consiglio24}. The good agreement between the ideal and the simulated results, despite the inherent errors in real quantum devices, strengthens the idea that 
{our model of a cyclic quantum battery} can be successfully 
{implemented} in current solid-state experimental platforms.

The paper is organized as follows. In Section \ref{Preliminaries} we discuss the basic aspects of the energy extraction from a  bipartite system. 
Section \ref{sec:theo_td_cycle} is devoted to the description of the thermodynamic cycle applied to the QB. 
In Section \ref{sec:min_model_squid} we introduce a model based on coupled superconducting qubits as an example of an experimentally feasible setup for the considered cyclic QB.
Section \ref{Results} is devoted to the numerical results concerning the ergotropy and the efficiency of the the cycle, which are compared with those obtained from the simulation carried out on IBM quantum machines. 
In Section \ref{Conclusions}, we present our main conclusions and outlooks. Details of the calculations and side remarks are deferred to various Appendices.

\section{Basics of ergotropy extraction}
\label{Preliminaries}

\subsection{Ergotropy and passive states}
Let $\varrho = \sum_{k=1}^d \lambda_k^\downarrow \vert \lambda_k^\downarrow \rangle\langle  \lambda_k^\downarrow \vert$ be the density matrix describing the state of a finite $d$-dimensional quantum system with eigenvalues sorted in  descending order, $\lambda_k^\downarrow \geq \lambda_{k+1}^\downarrow$, and let $H = \sum_{k=1}^d \epsilon_k^\uparrow \vert \epsilon_k^\uparrow \rangle\langle \epsilon_k^\uparrow \vert $ be its Hamiltonian with eigenvalues sorted in ascending order, $\epsilon_k^\uparrow \leq \epsilon_{k+1}^\uparrow$. Then, the average energy of this state can be decomposed as
\begin{equation}
E(\varrho) = \Tr[H\varrho] = \mathcal{E}(\varrho) + E(\pi),
\end{equation}
where $\mathcal{E}(\varrho)$ is the ergotropy of $\varrho$, i.e., the maximum amount of work that can be extracted via cyclic unitaries \cite{Hatsopoulos76, Allahverdyan_epl}. A cyclic unitary is generated by a time-dependent Hamiltonian which coincides at the initial and final times. The energy $E(\pi)=\Tr[H \pi]$, not extractable through such unitaries, is associated to the \textit{passive} state $\pi$ of $\varrho$, which is unitarily related to the latter via
\begin{equation}
\pi = U(\vec{\theta}) \varrho U^\dagger(\vec{\theta}) = \sum_{k=1}^d \lambda_k^\downarrow \vert \epsilon_k^\uparrow \rangle \langle \epsilon_k^\uparrow \vert,
\label{eq:theo_pi_pass_state}
\end{equation}
with 
\begin{equation}
U(\vec{\theta}) = \sum_{k=1}^{d} e^{i\theta_k}\vert \epsilon_k^\uparrow \rangle\langle \lambda_k^\downarrow \vert.
\label{eq:U_ergextr}
\end{equation}
The phases $\vec{\theta} = \{\theta_k\}$ reflect the arbitrariness in the normalization of the eigenstates, and only $d-1$ of them are 
relevant
(factoring out a global phase) \cite{Barra_njp, Satriani24}. Such phases are usually neglected because irrelevant for both the passive state \eqref{eq:theo_pi_pass_state} and the ergotropy
\begin{equation}
\mathcal{E}(\varrho) = E(\varrho)-E(\pi) = \sum_{j,k=1}^d \lambda_j^\downarrow \epsilon_k^\uparrow ( \vert \langle \lambda_j^\downarrow \vert \epsilon_k^\uparrow \rangle \vert^2 - \delta_{jk}),
\label{eq:theo_ergo}
\end{equation}
which is extracted from the system by the \textit{global} cyclic unitary \eqref{eq:U_ergextr}.
The ergotropy is non-negative, $\mathcal{E}(\varrho) \geq 0$, and zero if and only if $\varrho$ is passive, $\mathcal{E}(\pi) = 0$.  
Therefore, states with vanishing ergotropy, $\mathcal{E}=0$, are referred to as \textit{passive states} \cite{Pusz1978,Lenard1978}, and those with finite ergotropy, $\mathcal{E}>0$, as \textit{active states}.
The class of passive states comprises all states that commute with the system Hamiltonian and have no population inversions \cite{Pusz1978,Lenard1978,Allahverdyan_epl}. Therefore, thermal---or, equivalently, Gibbs---states of the form 
\begin{equation}
\tau_\beta = \frac{e^{-\beta H}}{\Tr[e^{-\beta H}]} = \sum_{k=1}^d \frac{e^{-\beta \varepsilon_k^\uparrow}}{\Tr[e^{-\beta H}]} \vert \epsilon_k^\uparrow \rangle \langle \epsilon_k^\uparrow \vert,
\label{eq:tau_gibbs}
\end{equation}
with $\beta = T^{-1}$ being the inverse temperature (throughout the paper we consider units in which $k_B=\hbar=1$), naturally belong to the class of passive states.  Within the latter class, thermal states maximize the entropy for a given energy and minimize the energy for a given entropy, thus they are the most stable.\footnote{For completeness, we mention that states that maximize the energy for a given entropy and also minimize the entropy for a given energy can also be identified \cite{Perarnau-Llobet_pre_2015}.}
Remarkably, work can be extracted from several independent copies of passive states by acting on them collectively, except for thermal states \cite{Alicki_2013,Niedenzu2019}. For this reason, a state $\varrho$, either thermal state or ground state (thermal state at $T=0$), is referred to as a \textit{completely passive} state, indicating that $\varrho^{\otimes n}$ is passive for all $n=1,2,\ldots$ with respect to the Hamiltonian $H^{(n)}=\sum_{k=1}^n H_k$, with $H_k$ the $k$-th independent copy of the Hamiltonian $H$ \cite{Pusz1978,Lenard1978}.

\subsection{Extracting ergotropy from the Gibbs state of a bipartite interacting system}
\label{sec:extr_ergo_from_multipart}

Work can be extracted by global cyclic unitaries acting on a collection of multiple copies of a passive state $\varrho$, provided it is not thermal. Accordingly, the resulting ergotropy is in general a non-extensive quantity, $\mathcal{E}(\varrho^{\otimes n} ) \geq n \mathcal{E}(\varrho)$, except when the passive state $\pi$ of $\varrho$ is thermal, in which case $\mathcal{E}(\varrho^{\otimes n} ) = n \mathcal{E}(\varrho)$ is extensive~\cite{Niedenzu2019}.

In this work, in a sense, we adopt a complementary approach: Since \textit{global} cyclic unitaries cannot extract work from the thermal state of a multipartite interacting system, we show that \textit{local} cyclic unitaries, acting separately on one of the subsystems after disconnecting the latter, can. This follows from the general observation that the reduced state of a subsystem of a globally thermal system can be active with respect to its local Hamiltonian, and hence it may contain extractable work when disconnected from the rest of the system. 
We stress that the presence of interactions is a necessary condition
for work extraction: If the multipartite system comprises $n$ non-interacting subsystems, then $H = \sum_{k=1}^n H_k$, with $H_k$ the Hamiltonian of the $k$-th subsystem, and the thermal state of the system is just the product state of the thermal states of the subsystems,
$\bigotimes_{k=1}^{n}\tau_\beta^{(k)}$.
So no local cyclic unitary process can extract work from the subsystems, each of which being in a thermal (passive) state.

For clarity, we restrict the discussion to a bipartite interacting system of the form
\begin{equation}
H(t) = H_A + H_B + \gamma(t) H_{\rm int} \equiv H_0 + \gamma(t) H_{\rm int},
\label{eq:H_A_B_int}
\end{equation}
where $H_{k}$ denotes the local Hamiltonian of the $k$-th subsystem---$k=A,B$---and $H_{\rm int}$ the interaction between the two. The coupling strength $\gamma(t)$ of the interaction is externally modulated in time and, for our purposes, we model it as a Heaviside step function, $\gamma(t) = \gamma_0 \Theta(t'-t)$ with $\gamma_0 \neq 0$ a dimensionless parameter so that the interaction is finite and constant before $t'$ and zero after $t'$, modeling an instantaneous switching off (disconnection of the two subsystems). 

The interacting bipartite system is initialized in the thermal state $\tau_\beta$ \eqref{eq:tau_gibbs} at a time $t_0 < t'$, $\gamma(t_0)=\gamma_0 \neq 0$. The interaction is then switched off at time $t'$, $\gamma(t')=0$, instantaneously disconnecting the two subsystems. After this operation, provided that the reduced states $\rho_{A(B)} = \Tr_{B(A)}(\tau_\beta)$ are not passive with respect to their local Hamiltonian $H_{A(B)}$, it is possible to extract ergotropy from the subsystems. For each subsystem, 
we define the local cyclic unitary $U_{k}(\vec{\theta}_k)$ \eqref{eq:U_ergextr} with $k=A,B$ and introduce
\begin{equation}
\mathcal{U}_l(\vec{\theta}_A,\vec{\theta}_B) = U_{A}(\vec{\theta}_A) \otimes U_{B}(\vec{\theta}_B).
\label{eq:calU_AB}
\end{equation}
The maximum work that can be extracted by a unitary of the form in Eq. \eqref{eq:calU_AB} is referred to as \textit{local} ergotropy \cite{Mukherjee_pre2016,Alimuddin_pra2019,Perarnau-Llobet_prx2015} and reads
\begin{eqnarray}
\mathcal{E}^{(l)}&\equiv\Tr\Big[(H_A + H_B)\Big(\varrho-\pi_l(\vec{\theta}_A,\vec{\theta}_B)\Big)\Big] \nonumber\\
&=\Tr[H_A (\rho_A-\pi_A)]+\Tr[H_B (\rho_B-\pi_B)]\nonumber\\
&=\mathcal{E}_A+\mathcal{E}_B \geq 0,
\label{eq:ErgLoc_theo}
\end{eqnarray}
where
\begin{equation}
\pi_l(\vec{\theta}_A,\vec{\theta}_B) =\mathcal{U}_l(\vec{\theta}_A,\vec{\theta}_B) \tau_\beta \mathcal{U}_l^\dagger (\vec{\theta}_A,\vec{\theta}_B).
\label{eq:theo_bip_after_ergo}
\end{equation}
Denoting by $\mathcal{E}^{(g)}$ the \textit{global} ergotropy extracted by a global unitary $\mathcal{U}_g$ acting on the overall system, the inequality $\mathcal{E}^{(l)} \leq \mathcal{E}^{(g)}$ holds, because the global unitary can extract work also from correlations, while the local unitaries only from the subsystems.

A few remarks are in order. Although the reduced state of the $k$-th subsystem is passive with respect to $H_k$, $\pi_{A(B)} = \Tr_{B(A)}[\pi_l(\vec{\theta}_A,\vec{\theta}_B)]$, the state \eqref{eq:theo_bip_after_ergo} is not the product state $\pi_l \neq \pi_A \otimes \pi_B$, in general, because  correlations 
are preserved by local unitary transformations. Moreover, the state $\pi_l(\vec{\theta}_A,\vec{\theta}_B)$ is not necessarily passive with respect to $H_A + H_B$, because ergotropy has been extracted locally, not globally. Indeed, the passive state of $\tau_\beta$ with respect to $H_A+H_B$ is $\pi_g = \mathcal{U}_g \tau_\beta \mathcal{U}_g^\dagger$. Furthermore, in contrast to global cyclic unitaries, for which phases are irrelevant for both ergotropy and final passive state, when ergotropy is extracted from subsystems, the coherences of the global state \eqref{eq:theo_bip_after_ergo} are crucially affected by such phases. 
To be more precise, as will be clearer in the following, the ergotropy extracted from each subsystem does not depend on the phases, and neither does their sum. However, if coherences and correlations between subsystems are not erased and further operations are performed on the system, then the result of the latter operations will depend, in general, on those phases. 

{Finally, we point out that the problem of optimal local work extraction from interacting subsystems has been recently addressed in \cite{Castellano24b}, giving rise to the concept of parallel ergotropy. The latter is non-negative and upper bounded by the global ergotropy by construction. Therefore, since the global ergotropy vanishes for a thermal state, we conclude that switching off the interaction between subsystems is a necessary condition to extract useful work from them when they are prepared in a globally thermal state.}

\section{Thermodynamic cycle of a bipartite quantum battery}
\label{sec:theo_td_cycle}

Given an interacting bipartite QB in a thermal state, weakly coupled with an external thermal reservoir, disconnecting the subsystems within the QB is what makes work extractable by cyclic unitary processes. In our scheme, the constant presence of a weakly coupled reservoir has a twofold motivation: First, it allows us to focus on the system Hamiltonian as the leading term, thus neglecting the explicit contribution of the reservoir during the connection/disconnection and work extraction strokes; Second, it makes unnecessary fully isolating the battery from the external environment, while still allowing for recharging by thermalization. Therefore, this assumption appears as well-suited to describe a realistic scenario.

We consider a bipartite system with Hamiltonian \eqref{eq:H_A_B_int}, where the external drive for the interaction, $\gamma(t)\in \{0,\gamma_0 \neq 0\}$, is a piecewise constant function which can be instantaneously switched off ($0$) or on ($\gamma_0$). The two subsystems forming the bipartite QB are said to be (dis)connected depending on the the presence (absence) of interaction between them, $\gamma(t)=\gamma_0$ ($\gamma(t)=0$).
In the following, we denote by $\varrho$ the state of the bipartite system and by $\rho_k$ the reduced state of the $k$-th subsystem, $k=A,B$. We assume to operate the working cycle while keeping the QB weakly coupled to a thermal bath at inverse temperature $\beta$.
{In addition,} we will assume all the four strokes of the cycle, except for the thermalization (last one), as instantaneous. This assumption is justified for weak system-reservoir
coupling and fast strokes (see below for concrete 
examples, in particular in the framework of superconducting quantum hardware).

Denoting by $\mathcal{E}^{(p)}$ the maximum work (ergotropy) extracted by the generic protocol $p$, we introduce the following protocols:

\begin{protocol}[$p=s$]
\label{prot:s}
The resource consists of a \textit{single} subsystem, say $A$. The extractable work $\mathcal{E}^{(s)} = \mathcal{E}_A$ is the ergotropy of the subsystem $A$. The process which accomplishes this task is the unitary
\begin{equation}
\mathcal{U}_s(\vec{\theta}_A) = U_{A}(\vec{\theta}_A) \otimes I_B.
\label{eq:calU_A}
\end{equation}
The main reason for processing just one subsystem 
is the interest in managing energy locally\footnote{We also 
mention as a possible reason the case in which the reduced state of subsystem $B$ is already passive with respect to $H_B$, $\Tr_A[\tau_\beta]=\pi_B$. This happens, e.g., for two interacting qubits with Hamiltonian  $H=\Omega_A \sigma_A^z + \Omega_B \sigma_B^z +\gamma_0 \sigma_A^z \otimes \sigma_B^z$---where $\Omega_B = 2 \Omega_A>0$, $\gamma_0 = 1.5 \Omega_A$, and $\beta = 2 / \Omega_A$---prepared in the thermal state $\tau_\beta$. The reduced state  $\rho_{A(B)} = \Tr_{B(A)}[\tau_\beta]$ is active (passive) with respect to the local Hamiltonian $H_{A(B)}$.}.
\end{protocol}

\begin{protocol}[$p=l$]
\label{prot:l}
The resource consists of both subsystems, $A$ and $B$, separately. The extractable work $\mathcal{E}^{(l)}=\mathcal{E}_A+\mathcal{E}_B$ is the \textit{local} ergotropy \eqref{eq:ErgLoc_theo}. The process which accomplishes this task is the local unitary $\mathcal{U}_l(\vec{\theta}_A,\vec{\theta}_B)$ in Eq. \eqref{eq:calU_AB}.
\end{protocol}

\begin{protocol}[$p=g$]
\label{prot:g}
The resource consists of both subsystems, $A$ and $B$, including correlations.
The extractable work $\mathcal{E}^{(g)}$ is the \textit{global} ergotropy \eqref{eq:theo_ergo}. The process which accomplishes this task is the global unitary $\mathcal{U}_g$ defined as in Eq. \eqref{eq:U_ergextr}. Note that this is possible because $\tau_\beta$  is completely passive with respect to the Hamiltonian \eqref{eq:H_A_B_int} with $\gamma(t)=\gamma_0$, but can be active with respect to $H_0=H_A+H_B$, which is the Hamiltonian of the bipartite system after disconnection. In this case, no arbitrary phases are included in the unitary, because irrelevant.
\end{protocol}

\begin{figure}[!tb]
	\centering
	\includegraphics[width=0.45\columnwidth]{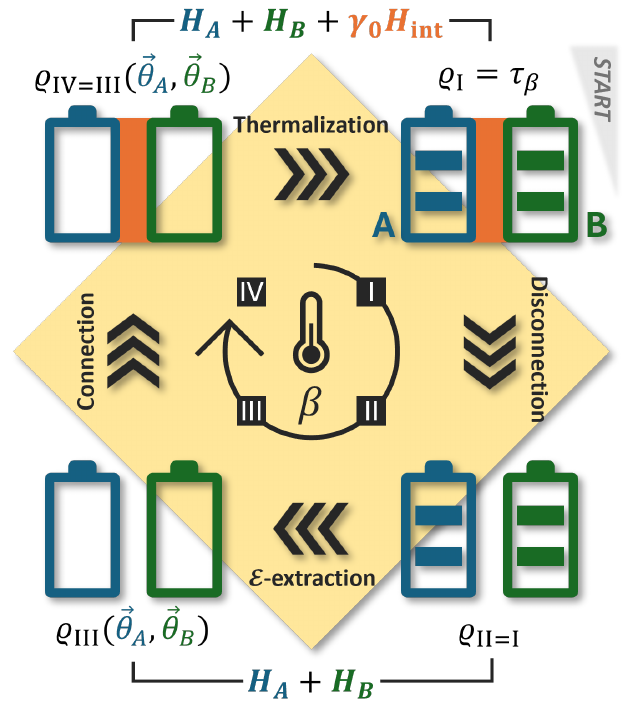}
	\caption{Working cycle of the bipartite cyclic QB in which both subsystems are the resource for extracting work [Protocol \ref{prot:l}]. The bipartite system is initially \textsc{(i)} prepared in a Gibbs state $\tau_\beta$ \eqref{eq:tau_gibbs} at inverse temperature $\beta$.
	\textsc{(i$\rightarrow$ii)} Disconnection of subsystems (switching off of $H_{\rm int}$);
	\textsc{(ii$\rightarrow$iii)} Local ergotropy extraction from each subsystem (pictorially represented as an empty battery since no further work can be extracted from it by a cyclic unitary) via the unitary $\mathcal{U}_l(\vec{\theta}_A,\vec{\theta}_B)$ \eqref{eq:calU_AB};
	\textsc{(iii$\rightarrow$iv)} Connection of subsystems (switching on of $H_{\rm int}$);
	\textsc{(iv$\rightarrow$i)} Thermalization to restore the initial Gibbs state and restart the cycle. Throughout the whole cycle, the bipartite system is 
    weakly coupled to a thermal reservoir at 
    inverse temperature $\beta$ (pictorially represented by the yellow shaded area).}
	\label{fig:td_cycle_AB}
\end{figure}

We now turn to the description of the four strokes of the working cycle of the cyclic QB, whose performance will depend on the protocol adopted to extract ergotropy. The cycle, e.g., for Protocol \ref{prot:l}, is sketched in Fig. \ref{fig:td_cycle_AB}. Starting 
from the bipartite system in a Gibbs state $\varrho_{\rm I}=\tau_\beta$ \eqref{eq:tau_gibbs} with $H=H_0 + \gamma_0 H_{\rm int}$, the following strokes are performed.

\textsc{(i$\rightarrow$ii)} \textit{Disconnection of subsystems.---}It is accomplished by switching off the interaction. After disconnection, the system is in the state $\varrho_{\rm II} = \varrho_{\rm I}$, since this instantaneous stroke leaves the state unchanged. Therefore, the energy cost for this stroke reads
\begin{equation}
E_d =-\gamma_0 \Tr[H_{\rm int}\varrho_{\rm I}],
\label{eq:Ed}
\end{equation}
and is independent of the protocol considered for extracting work. 

\textsc{(ii$\rightarrow$iii)} \textit{Extraction of ergotropy.---}It is accomplished by performing a cyclic unitary operation $\mathcal{U}_p$ on the state $\varrho_{\rm II}$, leading to $\varrho_{\rm III}^{(p)}(\theta) = \mathcal{U}_p(\theta) \varrho_{\rm II} \mathcal{U}_p^\dagger(\theta)$, where $p=s,l,g$ refers to protocols (i,ii,iii), respectively, and, for conciseness, $\theta$ denotes the possible arbitrary phases, if any. The maximum work thus extracted reads
\begin{equation}
\mathcal{E}^{(p)} = \Tr\left[(H_A+H_B) \left(\varrho_{\rm II} - \varrho_{\rm III}^{(p)}(\theta)\right)\right],
\label{eq:W}
\end{equation}
and does not depend on $\theta$. The amount of extracted work and the associated unitary, hence also the final state, depend on the chosen protocol:
(i) $\mathcal{E}^{(s)}=\mathcal{E}_A$ by $\mathcal{U}_s(\vec{\theta}_A)$ in Eq. \eqref{eq:calU_A};
(ii) $\mathcal{E}^{(l)}$ by $\mathcal{U}_l(\vec{\theta}_A,\vec{\theta}_B)$ in Eq. \eqref{eq:calU_AB};
(iii) $\mathcal{E}^{(g)}$ by $\mathcal{U}_g$ in Eq. \eqref{eq:U_ergextr}.
We point out that the final state $\varrho_{\rm III}^{(p=s,l)}(\theta)$ explicitly depends on the arbitrary phases, unlike the global passive state  $\varrho_{\rm III}^{(g)}$ \eqref{eq:theo_pi_pass_state}. The possible presence of arbitrary phases  will affect the result of the next stroke.

\textsc{(iii$\rightarrow$iv)} \textit{Connection of subsystems.---}It is accomplished by switching on the interaction. After connection, the system is in the state $\varrho_{\rm IV}^{(p)}(\theta) = \varrho_{\rm III}^{(p)}(\theta)$, since this instantaneous stroke leaves the state unchanged. Therefore, the energy cost for this stroke reads
\begin{equation}
E_c^{(p)}(\theta) = \gamma_0 \Tr[H_{\rm int}\varrho_{\rm III}^{(p)}(\theta)].
\label{eq:Ec}
\end{equation}

\textsc{(iv$\rightarrow$i)} \textit{Thermalization.---}The bipartite system is always in contact with an external thermal reservoir (weak coupling) throughout the whole cycle. During this last stroke, the system is left to naturally thermalize. This restores the initial thermal state at a vanishing energetic cost. In other words, this stroke can be intended as the process of recharging of the battery.

After characterizing this working cycle energetics, we define its efficiency as the ratio of the energy delivered by the battery to the energy cost to run the cycle \cite{Barra_njp, Hovhannisyan_prr}
\begin{equation}
\eta^{(p)}(\theta) = \frac{\mathcal{E}^{(p)}}{E_d+E_c^{(p)}(\theta)},
\label{eq:eta}
\end{equation}
with $0 \leq \eta^{(p)}(\theta) \leq 1$ and $p=s,l,g$, denoting the protocol. These bounds are explained as follows. The net result of the strokes \textsc{(i$\rightarrow\ldots\rightarrow$iv)} is a unitary evolution of the bipartite system, $\varrho_{\rm I} \to \varrho_{\rm IV}^{(p)}(\theta)$, and an associated cyclic change of the total Hamiltonian, $H^{\rm (I)}=H^{\rm (IV)}=H_0+\gamma_0 H_{\rm int}$ and $H^{\rm (II)} = H^{\rm (III)} = H_0$. Hence, the initial passive state can only increase its average energy, $\Tr\big[H(\varrho_{\rm IV}^{(p)}(\theta) - \varrho_{\rm I})\big] \geq 0$, and this increase must be equal to the non-negative work $E_d- \mathcal{E}^{(p)} + E_c^{(p)}(\theta) \geq 0$ performed on the system during the strokes \textsc{(i$\rightarrow \ldots \rightarrow$iv)}. The bounds on the efficiency follow accordingly, being $\mathcal{E}^{(p)} \geq 0$ by definition (ergotropies are non-negative quantities), and $E_d+ E_c^{(p)}(\theta) \geq \mathcal{E}^{(p)} $. We also note that our scheme differs from the one proposed in \cite{Barra_njp, Hovhannisyan_prr}, in which connection/disconnection of a quantum system from its environment was requested. 
{As a final remark here, we point out that protocols \ref{prot:s}--\ref{prot:g} are standard approaches to work extraction if compared, e.g., to the possibility of enhancing work extraction by feedback mechanisms \cite{Francica17,Morrone2023prapp,Satriani24}. Our purpose is to compare their different performances within a thermodynamic cycle for a quantum battery---i.e., not restricting the analysis to work extraction alone---without specifically inspecting the contribution to ergotropy arising from quantum coherence or correlations~\cite{Francica20,Francica_pre2022,Touil22}. However, the presence of the latter will be apparent in the fact that the efficiency depends on the arbitrary phases of the unitary under which work is extracted, $\eta^{(p)}(\theta)$, as well as in that $\mathcal{E}^{(l)} \leq \mathcal{E}^{(g)}$,  Protocol \ref{prot:g} having full access to coherences and correlations of the bipartite system.}

In the following, we theoretically describe a system based on superconducting circuits where the working cycle discussed above can be implemented in practice. Starting from that model, we compare the performance of the proposed model of cyclic QB depending on the adopted protocols \ref{prot:s}--\ref{prot:g} 
described above. After studying the ideal scenario, we simulate 
Protocol \ref{prot:s}---$p=s$, ergotropy extracted from a single subsystem---on an IBM quantum computer based on a superconducting qubit technology.

\section{Minimal model of cyclic bipartite quantum battery}
\label{sec:min_model_squid}
In this Section, we introduce and discuss a minimal example of the proposed model: A two-qubit cyclic QB. First, for illustrative purposes, we consider a simple 
feasible setup based on dc superconducting quantum interference device (SQUID) charge qubits which, under proper conditions, implements an effective model of two qubits with externally tunable interaction. Then, we elaborate on the thermodynamic cycle for the QB based on this system.

\subsection{Two-interacting-qubit model based on dc-SQUIDs}
\label{sec:squid}

\begin{figure}[!tb]
	\centering
	\includegraphics[width=0.7\columnwidth]{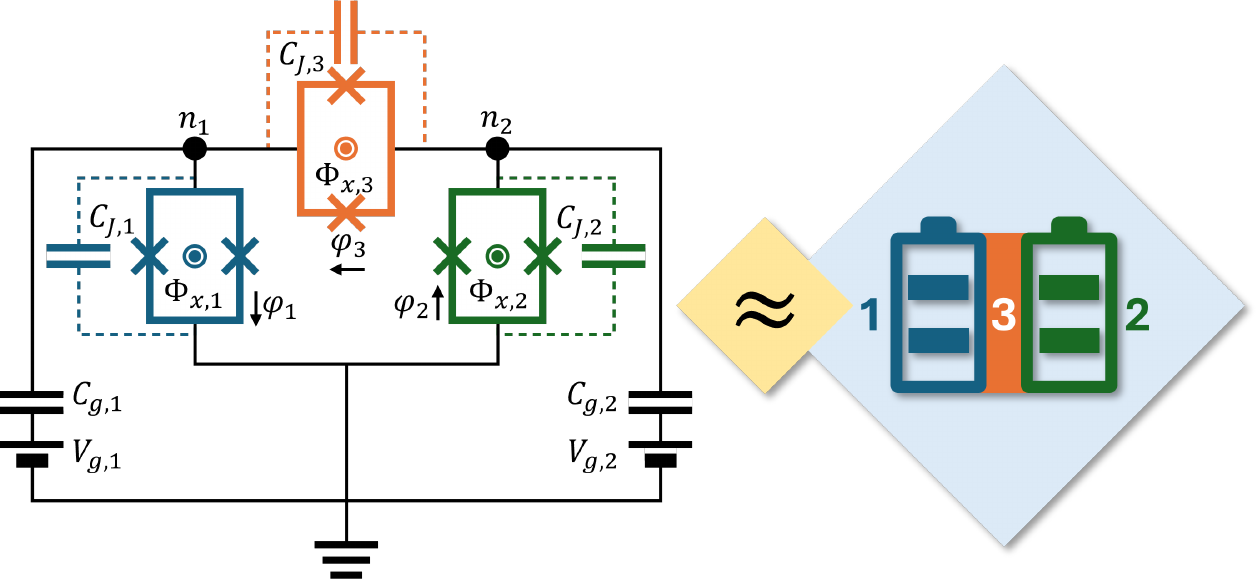}
	\caption{Schematic circuit of the proposed Josephson QB based on three dc-SQUIDs pierced by externally tunable magnetic fluxes. Josephson junctions are denoted by a cross and the black circles, labelled by $n_1$ and $n_2$, correspond to the first and second Cooper pair box, respectively. 
	The three dc-SQUIDs operate under the qubit approximation. The qubit 3, used as a quantum coupler, mediates an effective interaction between qubits 1 and 2. For the present setup to be used as a cyclic quantum battery we set $C_{J,3}=0$, so that $E_3=0$ in Eq. \eqref{eq:Ham_dcsquid}. Removing the flux-independent interaction between qubits 1 and 2 allows them to be fully disconnected by the externally-driven magnetic flux.}
	\label{fig:dcsquids}
\end{figure}

We consider a 
setup based on three symmetrical dc-SQUIDs (Fig. \ref{fig:dcsquids}), originally proposed in \cite{Wang_prb} as quantum storage unit and later as a possible experimental setup for a Maxwell’s demon-assisted quantum heat engine  \cite{Quan2006}.
The  Hamiltonian of the system comprises 
charging (Coulomb) and Josephson energy contribution.
When the Josephson energy 
is much weaker than the charging energy the evolution is approximately confined into the two-dimensional space spanned by the eigenstates of the charge operator, $\left\{ \left\vert 0\right\rangle_{k},\left\vert 1\right\rangle _{k}\right\} $ of the $k$-th Cooper pair box ($k=1,2,3$). Under the latter condition and the constraint $\varphi_{1}+\varphi_{2}+\varphi_{3}=0$ on the phase differences $\varphi_k$ in the $k$-th SQUID, it is possible to derive an effective (pseudo)-spin Hamiltonian of the form
\begin{eqnarray}
H =&\sum_{k=1}^{2}\Omega _{k}\sigma_{k}^{z}+E_{3}\sigma_{1}^{z}\sigma_{2}^{z}
-\sum_{k=1}^{2}E_{J,k}\cos\left( \pi \frac{\Phi _{x,k}}{\Phi_{0}}\right)\sigma_{k}^{x} \nonumber \\
&-E_{J,3}\cos \left( \pi \frac{\Phi _{x,3}}{\Phi _{0}}\right) \left(\sigma_{1}^{x}\sigma_{2}^{x}-\sigma_{1}^{y}\sigma_{2}^{y}\right).
\label{eq:Ham_dcsquid}
\end{eqnarray}
In this setup, the bipartite QB comprises the first two SQUIDs, while the third one plays the role of a tunable quantum coupler between them~\cite{Sete_prappl, Campbell23, Heunisch23}. In the Hamiltonian \eqref{eq:Ham_dcsquid} $E_{J,k}$ and $\Phi _{x,k}$ are, respectively, the Josephson energy and the magnetic flux in the $k$-th SQUID, while $\Phi _{0}=h/2e$ is the superconducting flux quantum, with $h$ the Planck constant and $e$ the elementary charge. Moreover, $\Omega _{k}= E_{C,k}\left( n_{g,k}-1/2\right) +2E_{3}\left( n_{g,j}-1/2\right)$, with $j\neq k$, where $n_{g,k}=C_{g,k}V_{g,k}/2e$, with $V_{g,k}$ and $C_{g,k}$ being respectively the gate voltage and capacitance in the $k$-th Cooper pair box.
The other coefficients are
$E_{C,k}=2e^{2}C_{\Sigma, j}/\left( C_{\Sigma, 1}C_{\Sigma, 2}-C_{J,3}^{2}\right) $, with $j \neq k$,
and $E_{3}=e^{2}C_{J,3}/2 \left( C_{\Sigma,1}C_{\Sigma, 2}-C_{J,3}^{2}\right) $,
with $C_{\Sigma,k}=C_{J,k}+C_{J,3}+C_{g,k}$. In Eq. \eqref{eq:Ham_dcsquid} the Pauli matrices are represented in the bases $\{\vert 0 \rangle_k,\vert 1 \rangle_k\}$ spanned by the eigenstates of the number operator of Cooper pair on the $k$-th box with zero and one Cooper pair, respectively, and read $\sigma _{k}^{x}= \vert 1\rangle _{kk}\langle 0 \vert + \vert 0\rangle_{kk}\langle 1\vert$, $\sigma _{k}^{y}= i(\vert 1\rangle _{kk}\langle 0 \vert - \vert 0\rangle _{kk}\langle 1 \vert)$ and $\sigma _{k}^{z}=\vert 0\rangle _{kk}\langle 0 \vert - \vert 1\rangle _{kk}\langle 1 \vert $.

{One of the key advantages of the proposed basic scheme (Fig. \ref{fig:dcsquids}), which relies on dc-SQUID charge qubits \cite{Makhlin1999,Nakamura1999} designed to operate in the regime $k_B T \ll E_C$~\cite{Clarke2008} and $E_C \gg E_J$, is its experimental feasibility. Therefore, we now examine experimentally feasible parameters considering comparable platforms and setups. Suitable values of the junction capacitance are in the range of femtofarad and below, $C_{J,k} \leq 1$ fF, while the gate capacitances $C_{g,k}$ can be even smaller \cite{Mahklin2001} (e.g., $C_{g,k}\sim 0.6$ aF and 
$C_{J,k}\sim 620$ aF \cite{Pashkin_2003}). 
The dimensionless gate charge usually takes values $n_{g, k} \sim 0 \div 2$ \cite{Vion_2002,Guthrie_2022}.
Charging energies $E_{C,k}/h \sim 10 \div 100$ GHz \cite{Pashkin_2003} and Josephson energies $E_{J,k}/h \sim 1 \div 10$ GHz fulfilling the condition $E_C \gg E_J$ are consistent with experimental setups \cite{Pashkin_2003}.
Regarding the dc-SQUID geometry, an effective area can be estimated in $A \approx  4.9$ $\mu {\rm m}^2$ which, provided a magnetic field $B_x \sim 0.1\div1$ mT, results in a magnetic flux threading the SQUID loop $\Phi_x / \Phi_0 = B_x A /\Phi_0 \sim 0.237 \div 2.37 $, with $\Phi_0 = 2.067$ fWb~\cite{Strambini_2020}. Assuming $n_{g, k} = 1$\footnote{
{For later convenience, we require $n_{g,k}=n_g > 1/2$ to ensure $\Omega_k=\Omega > 0$ in Eq. \eqref{eq:Ham}}.} and $C_{g,k} = 2$ aF, we can estimate the gate voltage $V_{g, k} = 2e n_{g,k}/C_{g,k} \approx 160$ mV.
Finally, a typical temperature is $T \sim 30\,\rm{mK}$ ($k_B T/h \sim 0.625\,\rm{GHz}$) \cite{Nakamura1999}.}

\subsection{Cyclic Josephson quantum battery}
\label{sec:rjqb}

\subsubsection{Hamiltonian}
Based on the dc SQUIDs setup described in Sec. \ref{sec:squid}, we introduce a Josephson QB (JQB) with Hamiltonian
\begin{equation}
H =  \Omega \sum_{k=1}^{2} \sigma_k^{z}  - E_{J}\cos\left( \pi \frac{\Phi _{x}}{\Phi_{0}}\right)
\Bigg[\sum_{k=1}^{2} \sigma_{k}^{x} + (\sigma_1^x\sigma_2^x - \sigma_1^y\sigma_2^y)\Bigg],
\label{eq:Ham}
\end{equation}
obtained from Eq.~(\ref{eq:Ham_dcsquid}) under the following assumptions: $E_3=0$; $\Phi_{x,k} = \Phi_{x}$ and $E_{J,k}=E_J$ for $k=1,2,3$; $E_{C,k}=E_C$ and $n_{g,k}=n_g$, from which $\Omega_k=\Omega = E_{C}\left( n_{g}-1/2\right)$, for $k=1,2$. In addition, the two-level approximation requires $E_{C} \gg E_{J}$.
The assumption of equal magnetic fluxes $\Phi_x$ is motivated by the will of reducing the number of free parameters, together with the fact that, considering SQUIDs with the same loop geometry, it is experimentally easier to  have an approximately homogeneous magnetic field piercing the three Cooper pair boxes, instead of strong inhomogeneities at  nanoscale~\cite{Liu_2005}.
Letting $\Phi_x = \Phi_x(t)$, the magnetic flux can be regarded to as the external driving which modulates the interaction 
$\sigma_1^x\sigma_2^x - \sigma_1^y\sigma_2^y$---in particular, switches it on or off---but also the single-qubit contributions $\propto \sigma_k^x$, a side effect specific of the model from which we derived the Hamiltonian. Switching on/off the interaction, however, in general 
does not suffice to fully disconnect the two qubits, because of the residual interaction $\propto \sigma_1^z\sigma_2^z$, independent of the magnetic flux (see Eq. \eqref{eq:Ham_dcsquid}). The assumption $E_3=0$, accomplished by setting $C_{J,3}=0$, allows us to discard the interaction $\propto \sigma_1^z\sigma_2^z$ fully disconnecting qubits through the magnetic flux.

Considering $\Omega$ as the reference scale for the energies, we can further simplify the Hamiltonian into
\begin{eqnarray}
H(t)
&= \sigma_A^{z}+\sigma_B^{z}-\gamma(t) (\sigma_A^{x} + \sigma_B^{x} + \sigma_A^x\sigma_B^x - \sigma_A^y\sigma_B^y)\nonumber\\
&\equiv H_{0} + \gamma(t) H_{\rm int},
\label{eq:Ham_gamma}
\end{eqnarray}
where, for later convenience, we have identified the bare Hamiltonian, $H_0$, and the driven one, $H_{\rm int}$ with $\gamma(t) \in \{0,\gamma_0 \neq 0 \}$. Notice that $H_{\rm int}$ includes single-qubit terms, in addition to the two-qubit interaction.
(Dis)connecting the two qubits is related to switching on (off) $H_{\rm int}$. 
{In the dc-SQUIDs system, after identifying $\gamma(t) \equiv E_{J}\cos\left( \pi \Phi _{x}(t)/\Phi_{0}\right)/\Omega$, this is accomplished by tuning the magnetic flux to $\Phi_{x}^{\rm (off)} = \left(m+1/2\right) \Phi_0$, with $m \in \mathbb{Z}$, for full disconnection ($\gamma(t) = 0$) and to any other value, $\Phi_{x}^{\rm (on)} \neq \Phi_{x}^{\rm (off)}$, for a connection at arbitrary coupling strength ($\gamma(t)=\gamma_0\equiv E_{J}\cos ( \pi \Phi _{x}^{\rm (on)}/\Phi_{0})/\Omega$). Therefore, for a given configuration of the setup, $\gamma_0$ can be externally tuned through the magnetic flux without changing the circuit elements and can take negative values depending on the value of the cosine function. Its possible values are limited to $\vert \gamma_0 \vert \ll \left(n_g - 1/2\right)^{-1}$ due to the condition $E_C \gg E_J$.} 

\subsubsection{Energetics of the working cycle of the JQB}
\label{sec:energetics_wc_JQB}
We study here the cycle described in Sec. \ref{sec:theo_td_cycle} for the JQB, 
for all the three protocols $p=s,l,g$ for extracting work. We recall that the external driving of our model, ultimately the magnetic flux, not only affects the interaction between qubits but also the local Hamiltonian of each qubit. To simplify the notation, we denote by $\langle O \rangle \equiv \Tr[O \varrho_{\rm I}]$ the expectation value of the generic observable $O$ on the initial Gibbs state $\varrho_{\rm I }=\tau_\beta$ and we omit the $\otimes$ symbol and identities when clear from the context. We also recall that the two-qubit state is denoted by $\varrho$ and the reduced state of the $k$-th qubit by $\rho_k$, with $k=A,B$.

First, we discuss the extraction of ergotropy from one qubit, as it is the building block for the working cycles with local unitaries. After the instantaneous disconnection, the local Hamiltonian of the $k$-th qubit is just $H_k = \sigma_k^z$. The reduced state of the qubit is $\rho_{A(B),{\rm II}} = \Tr_{B(A)}[\tau_\beta]$. Notice that, since the two qubits are identical and the thermal state $\tau_\beta$ is symmetric under exchange of the two qubits (symmetry inherited from the Hamiltonian \eqref{eq:Ham_gamma}), the reduced states of the qubits are equal, $\rho_{k,{\rm II}} \equiv \rho_{\rm II}$. A qubit state can be represented in the Bloch sphere as a point of coordinates $(x,y,z)$, where $a = \Tr[\sigma_a \rho_{\rm II}]$ with $a=x,y,z,$ and $r = \sqrt{x^2+y^2+z^2}$ (see \ref{app:eig_rho_qubit} for more details). To extract ergotropy from the battery, we apply the unitary operation (see \ref{app:U_erg})
\begin{equation}
U(\theta) =
\left(
\begin{array}{cc}
 e^{i \theta}  \cos\alpha & e^{i \theta}  \sin\alpha \\
-e^{-i \theta} \sin\alpha & e^{-i \theta} \cos\alpha
\end{array}
\right)
=e^{i \theta \sigma_z}e^{i \alpha \sigma_y},
\label{eq:Uerg_qiskit}
\end{equation}
where $0 \leq \theta < \pi$ is an independent arbitrary phase and $\alpha = \arctan[-(r+z)/x]$, with $r = \sqrt{x^2+z^2}$, is determined by the qubit state. Notice that here we have explicitly taken into account that $y=0$ due to the fact that the Gibbs state, and so $\rho_{\rm II}$, is a real density matrix. The passive state of the qubit with respect to $\sigma_z$, the spectrum of the latter being non-degenerate, is unique $\rho_{\rm III}= U(\theta) \rho_{\rm II}U^\dagger(\theta) = \frac{1}{2}(1-r\sigma_z)$, independent of $\theta$, and thermal---being it a two-level system \cite{Allahverdyan_epl}---at an effective inverse temperature $\beta_{\rm eff} = \rm{arctanh}(r)$. A unitary transformation preserves the purity of a state, thus both $\rho_{\rm II}$ and $\rho_{\rm III}$ have the same $r$ (see \ref{app:eig_rho_qubit}).
From this qubit, the ergotropy extracted by the unitary \eqref{eq:Uerg_qiskit} is
\begin{equation}
\mathcal{E}  
= z+r 
=  \langle\sigma_A^z\rangle + \sqrt{\langle\sigma_A^x\rangle^2 + \langle\sigma_A^z\rangle^2}.
\label{eq:ideal_Erg} 
\end{equation}

Now, we analytically inspect the energetics of the strokes in the working cycle based on the extraction of the ergotropy $\mathcal{E}^{(s)}$ from the single qubit [Protocol \ref{prot:s}], of the local ergotropy $\mathcal{E}^{(l)}$ from both the qubits [Protocol \ref{prot:l}], and of the global ergotropy $\mathcal{E}^{(g)}$ [Protocol \ref{prot:g}] from both the qubits, including correlations. Given the thermal state $\tau_\beta$, we recall that work is extracted with respect to $H_0$. Deferring all the details to \ref{app:ideal_energies}, here we quote the final results.
For all three considered scenarios, the energy cost for disconnection is
\begin{equation}
E_d = \gamma_0 ( 2\langle\sigma_A^x\rangle + \langle\sigma_A^x\sigma_B^x\rangle - \langle\sigma_A^y\sigma_B^y\rangle).
\label{eq:ideal_Ed}
\end{equation}
The work, instead, differs:
$\mathcal{E}^{(s)}=\mathcal{E}$ \eqref{eq:ideal_Erg} extracted by $\mathcal{U}_s(\theta)$ \eqref{eq:calU_A}, and 
$\mathcal{E}^{(l)}=2\mathcal{E}$ by  $\mathcal{U}_l(\theta_A,\theta_B)$ \eqref{eq:calU_AB}, where the single-qubit unitary is given in Eq. \eqref{eq:Uerg_qiskit}.
The global ergotropy is extracted by $\mathcal{U}_g$, as per definition \eqref{eq:U_ergextr}, and reads
\begin{equation}
    \mathcal{E}^{(g)} = 2  \langle \sigma_A^z \rangle
    + \frac{2}{Z}\left( e^{-\beta \epsilon_1^\uparrow} - e^{-\beta \epsilon_4^\uparrow} \right) 
    \geq \mathcal{E}^{(l)},
    \label{eq:ideal_Ergo_global}
\end{equation}
where $Z = \Tr[e^{-\beta H}]=\sum_{k=1}^{4}e^{-\beta \epsilon_k^\uparrow}$ is the partition function and $\epsilon_1^\uparrow$ and $\epsilon_4^\uparrow$ are, respectively, the lowest and the highest energy level of $H$ \eqref{eq:Ham_gamma} with $\gamma(t) = \gamma_0$ (see \ref{app:Hspectrum} for their analytical expression).
The state at the end of the stroke \textsc{(ii$\rightarrow$iii)} clearly depends on the chosen protocol, and so does the energy cost for connection
\begin{eqnarray}
E_c^{(s)}(\theta) =& -\gamma_0 \Big\{ \langle\sigma_A^x\rangle + \cos(2\theta)\Big[\cos(2\alpha)(\langle\sigma_A^x\rangle + \langle\sigma_A^x\sigma_B^x\rangle)\nonumber\\
&-\langle\sigma_A^y\sigma_B^y\rangle - \sin(2\alpha)(\langle\sigma_A^z\rangle + \langle\sigma_A^z\sigma_B^x\rangle) \Big] \Big\}, \label{eq:ideal_Ec}
\end{eqnarray}
where $\theta$ is the arbitrary phase of the unitary \eqref{eq:Uerg_qiskit} acting on the single qubit,
\begin{eqnarray}
E_c^{(l)}(\theta) =& -\gamma_0 \cos(2\theta) \Big(\cos^2(2\alpha) \langle\sigma_A^x\sigma_B^x\rangle -\langle\sigma_A^y\sigma_B^y\rangle \nonumber\\
& +\sin^2(2\alpha)\langle\sigma_A^z\sigma_B^z\rangle  - \sin(4\alpha) \langle \sigma_A^x \sigma_B^z\rangle \Big),
\label{eq:ideal_Ec_twoqb}
\end{eqnarray}
where $\theta \equiv \theta_A+\theta_B$ is the sum of the arbitrary phases $\theta_k$ of each local unitary $U_k(\theta_k)$ acting on the $k$-th qubit, and $E_c^{(g)} \equiv 0$. Note that in the $l$-JQB, although the local ergotropy is separately extracted from each qubit, there is only one independent effective parameter.

As a result, the efficiency of the cycle $\eta^{(p)}(\theta) = \mathcal{E}^{(p)}/ [ E_d + E_c^{(p)}(\theta)]$, with $p=s,l$,  takes distinct values only for $0 \leq \theta \leq \pi/2$, and its extreme values are attained at $\theta = 0,\pi/2$. The exact correspondence between $\theta=0,\pi/2$ and the minimum or maximum value of the efficiency depends on the sign of the quantity within brackets, multiplying $\cos(2\theta)$ in \eqref{eq:ideal_Ec}
and \eqref{eq:ideal_Ec_twoqb}. Instead, in the case of global extraction of ergotropy, the efficiency reads $\eta^{(g)} = \mathcal{E}^{(g)}/E_d$.


\section{Results}
\label{Results}
In this Section, we investigate the performance of the proposed JQB described by the Hamiltonian \eqref{eq:Ham_gamma}. First, we discuss the ideal performance of the working cycle of the $p$-JQB, then we simulate the case of the $s$-JQB on actual IBM superconducting quantum machines. 

\begin{figure*}[!tb]
\includegraphics[width=0.9\textwidth]{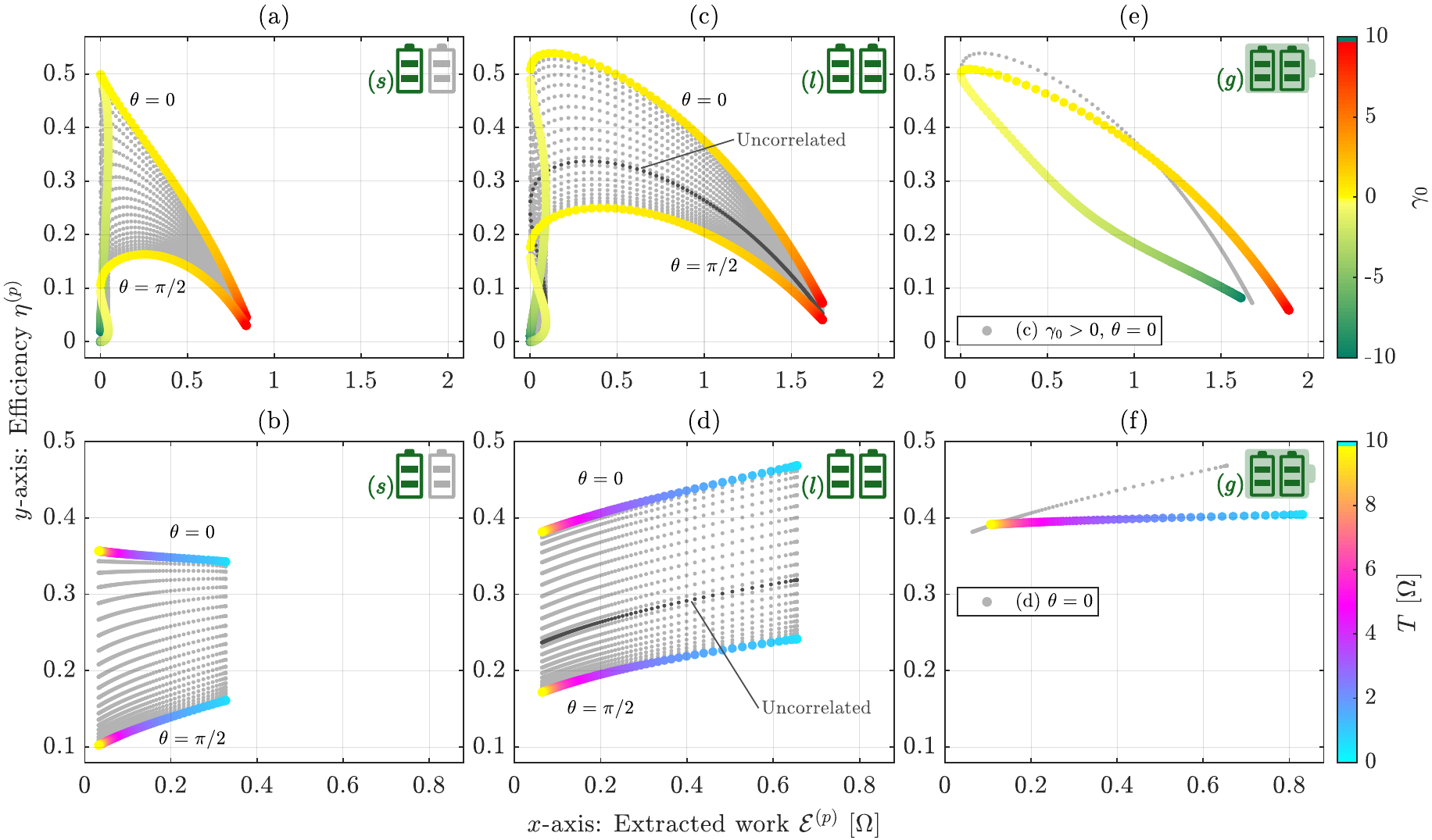}
\caption{Performance of the $p$-JQB operated under different protocols $p$ in terms of efficiency $\eta^{(p)}$ vs extracted work $\mathcal{E}^{(p)}$: (a,b) $s$-JQB (single ergotropy, one qubit as a resource); (c,d) $l$-JQB (local ergotropy, both qubits as a resource); and (e,f) $g$-JQB (global ergotropy, both qubits and correlations as a resource). (a,c,e) Parametric plot of $\eta^{(p)}$ vs $\mathcal{E}^{(p)}$ for $0 \leq \theta \leq \pi/2$ and $-10 \leq \gamma_0 \leq 10$ at $T = 0.5\Omega$. (b,d,f) Same as in (a,c,e) but for $0 \leq T \leq 10\Omega$ at $\gamma_0 = 0.8$. (a-d) Optimal tradeoff between $\eta$ and $\mathcal{E}$ (Pareto front) is obtained at $\theta = 0$ with $\gamma_0 > 0$.
(e,f) Results do not depend on arbitrary phases, since ergotropy is extracted by a global cyclic unitary. In panels (c) and (d), as references, we have reported the results for the case in which disconnection also erases correlations between the two qubits (dark gray dots). In panels (e) and (f), as references, we have reported the best results achieved in panels (c) and (d) respectively (gray dots). As a benchmark for the ergotropy, we
{indicate here} that the energy-level spacing of the qubit is $2\Omega$
{---see  \ref{app:U_erg}}. For sake of completeness, we have considered values of $\gamma_0$ and of $T$ in a range wider than the typical values admissible for the specific superconducting setup discussed in Sec. \ref{sec:squid}.}
\label{fig:idperf_LG_qbbat}
\end{figure*}

\subsection{Ideal performance}
\label{sec:id_perf}
The energetics of the working cycle of the $p$-JQB, operated under the protocol $p$ for extracting work, is summarized by the $(\mathcal{E}^{(p)},E_d,E_c^{(p)}(\theta))$, which refers to the extracted work and the energy cost of disconnection and connection, respectively. Here, we compare the performance of the following three working protocols $p=s,l,g$ introduced and described in Sec. \ref{sec:theo_td_cycle}:
\begin{enumerate}[(i)]
\item $s$-JQB: $(\mathcal{E}^{(s)}=\mathcal{E},E_d,E_c^{(s)}(\theta))$, with extraction of ergotropy from a single qubit;
\item $l$-JQB: $(\mathcal{E}^{(l)}=2\mathcal{E},E_d,E_c^{(l)}(\theta))$, with local extraction of ergotropy from both qubits;
\item $g$-JQB: $(\mathcal{E}^{(g)},E_d,E_c^{(g)}=0)$, with global extraction of ergotropy from both qubits, including correlations;
\end{enumerate}
where $\mathcal{E}$ is the ergotropy \eqref{eq:ideal_Erg} extracted from one qubit, $\mathcal{E}^{(g)}$ is the global ergotropy \eqref{eq:ideal_Ergo_global}, $E_d$---the energy of disconnection \eqref{eq:Ed}---is the same in all the three protocols, and $E_c^{(s,l)}(\theta)$ is the energy of connection of the $(s,l)$-JQB, see Eqs. \eqref{eq:ideal_Ec}--\eqref{eq:ideal_Ec_twoqb}.

The performance of the three JQBs described above is summarized in Fig. \ref{fig:idperf_LG_qbbat}, where the extracted work $\mathcal{E}^{(p)}$ with corresponding efficiency of the cycle $\eta^{(p)}(\theta)$ is shown varying $\gamma_0$ at fixed temperature $T=0.5\Omega$ (upper panels) or varying $T$ at fixed $\gamma_0 = 0.8$ (lower panels). First, the efficiency of the $(s,l)$-JQB is maximum for $\theta = 0$, and minimum for $\theta = \pi/2$ [panels (a-d)]. Looking for the optimal trade-off (Pareto front) between $\mathcal{E}^{(p)}$ and $\eta^{(p)}(\theta)$, i.e., the conditions $(\gamma_0,\theta)$ under which we cannot further improve one quantity without degrading the other, we conclude that the optimal performance is obtained for $\theta = 0$ with $\gamma_0>0$. Second, $\mathcal{E}^{(l)}=2\mathcal{E}^{(s)}=2\mathcal{E}$, trivially, but, remarkably, the local extraction of ergotropy from both qubits also enhances the efficiency [panels (a-d)]. Third, the most striking result is the presence of parameter regions where, at parity of extracted work, the $l$-JQB is more efficient than the $g$-JQB [panels (c-f)]. A few remarks are in order to clarify this point. The protocols considered in this working cycle, i.e., the unitaries in the stroke \textsc{(ii$\rightarrow$iii)}, aim at maximizing the work extractable by cyclic unitary processes (ergotropy), not the efficiency of the cycle. Then, a constrained optimization of the efficiency is performed with respect to the arbitrary phases of the unitary, if any. The constraint is that the unitary must extract ergotropy and, for the $g$-JQB, phases are irrelevant. When the optimization target is the ergotropy, which is the present case, clearly $\mathcal{E}^{(g)} \geq \mathcal{E}^{(l)}$. Limited to the regime in which both the $l$- and the $g$-JQB extract the same amount of work, there exist a set of parameters under which $\eta^{(l)} > \eta^{(g)}$. 

There is another point to stress in order to get an intuitive reason of why, provided the same extracted work, the efficiency of the $l$-JQB can exceed that of the $g$-JQB.\footnote{Note that, in general, at a given temperature the $l$-JQB provides the same ergotropy as that of the $g$-JQB for different values of $\gamma_0$, which, in turn, also results in different values of $E_d$ \eqref{eq:ideal_Ed}.} One could naively expect $\eta^{(g)} \geq \eta^{(l)}(\theta)$, because of $E_c^{(g)}=0$. However, the fact is that $E_c^{(l)}(\theta)$ is not necessarily an energy cost, positive, because, as shown in Eq. \eqref{eq:ideal_Ec_twoqb}, the magnitude and sign of the latter depend on $\theta$ as $E_c^{(l)}(\theta) \propto \cos(2 \theta)$. Whenever the chosen parameter regions for the $l$- and $g$-JQB allow for the same ergotropy extraction and the total energy cost for disconnection/connection in the $l$-JQB is lower than the energy cost of disconnection in the $g$-JQB, we observe that $\eta^{(l)} > \eta^{(g)}$ [panel (e) of Fig.~\ref{fig:idperf_LG_qbbat}] -- see also Eq. \eqref{eq:eta}.

Concerning the role of temperature at given $\gamma_0=0.8$ [Fig. \ref{fig:idperf_LG_qbbat} (b,d,f)], we observe that the lower the temperature, the higher the extracted work. The dependence of the efficiency on the $T$ is weak or even negligible in the $s$- and in the $g$-JQB, respectively, while it is more pronounced in the $l$-JQB. For the $l$-JQB, higher (lower) values of ergotropy and efficiency are attained at lower (higher) $T$.  

Both the $l$-JQB and the $g$-JQB extract work from both the qubits, but the global unitary in the $g$-JQB has also access to the correlations between qubits. This additional resource for ergotropy extraction results in $\mathcal{E}^{(g)}\geq\mathcal{E}^{(l)}$. This concept is nicely captured by the notion of ergotropic gap, $\Delta_{\rm EG} \equiv \mathcal{E}^{(g)}-\mathcal{E}^{(l)}$. This quantity is always  nonvanishing in the presence of quantum correlations \cite{Mukherjee_pre2016,Alimuddin_pra2019,Perarnau-Llobet_prx2015}. Vice versa, instead, a nonzero ergotropic gap does not necessarily imply the presence of quantum correlations. Restricting ourselves to the entanglement of formation, we have explored how the latter depends on the parameters of the model, showing the beneficial role of the coupling strength 
$\gamma_0$ and the detrimental one of the temperature $T$ when they increase. In the $s$-JQB, ergotropy can be extracted also at high temperature, in the absence of entanglement, and the behavior of the ergotropic gap is consistent with the expected one. For further details, please refer to \ref{app:entg_erggap}.

Focusing on the $l$-JQB, we can consider the case in which the instantaneous disconnection also erases the correlations between the qubits, i.e., $\varrho_{\rm II} = \rho_A \otimes \rho_B$ with $\rho_A = \Tr_A[\tau_\beta]=\rho_B$ (because of the symmetry of $\tau_\beta$). The energy cost of disconnection $E_d=\Tr[(H_A+H_B)\rho_A \otimes \rho_B]-\Tr[H \tau_\beta]$ turns out to be equal to Eq. \eqref{eq:ideal_Ed}. By construction, the ergotropy is the same as $\mathcal{E}^{(l)}$. After ergotropy extraction, the system is in the state $\varrho_{\rm III} = \frac{1}{4}(1-r\sigma_A^z) \otimes (1-r\sigma_B^z)$, which is diagonal in the computational basis, $\{\vert 00 \rangle,\vert 01 \rangle,\vert 10 \rangle,\vert 11 \rangle\}$,
and so the energy cost of connection is zero (the proof is analogous to that of $E_c^{(g)}$ in \ref{app:null_Ec_glob}). Therefore, the efficiency of this cycle reads $\eta^{(l)}_{\rm uncorr}= \mathcal{E}^{(l)}/E_d = \eta^{(l)}(\pi/4)$, since $E_c^{(l)}(\pi/4)=0$, see Eq. \eqref{eq:ideal_Ec_twoqb}. This result highlights the relevance of phase coherence in the presence of correlations in our model. 
Indeed, as shown in Fig. \ref{fig:idperf_LG_qbbat}(c,d),
with a suitable choice of the angle $\theta$ 
the performance of the correlated $l$-JQB can be consistently improved with respect to the uncorrelated case. The latter, instead, cannot benefit from phase-coherence because once correlations are erased, the state after ergotropy extraction is independent of the phases introduced by the unitary \eqref{eq:U_ergextr}.

\begin{figure}[!tb]
\centering
\includegraphics[width=0.7\columnwidth]{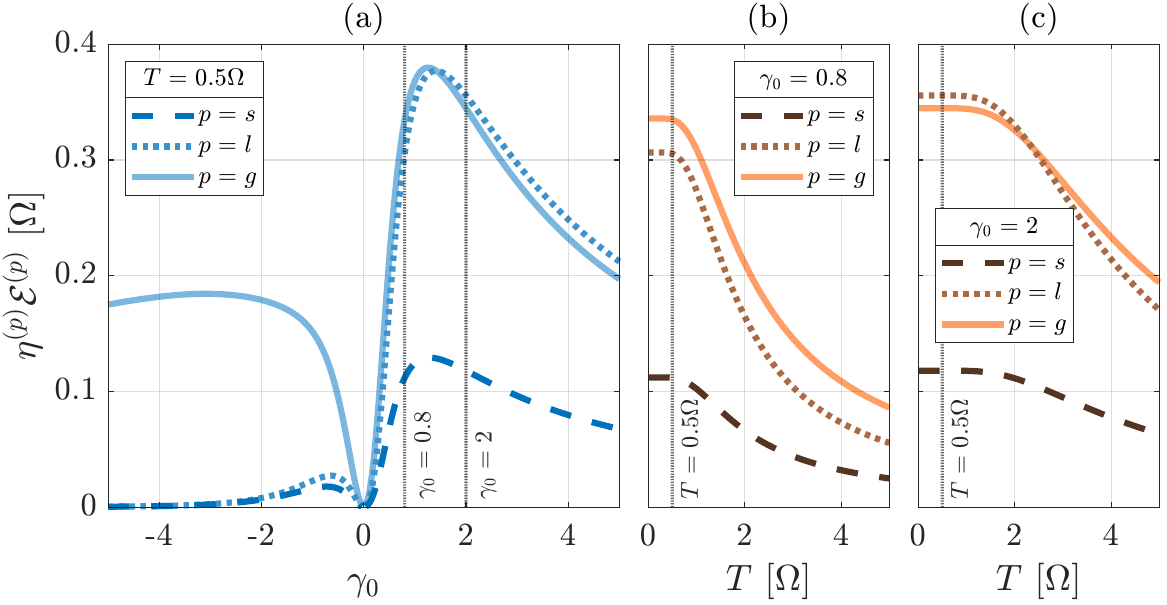}
\caption{
{Efficient extracted work $\eta^{(p)}\mathcal{E}^{(p)}$ (a) as a function of $\gamma_0$ at $T=0.5\Omega$ [see also Fig. \ref{fig:idperf_LG_qbbat}(a,c,e)] and as a function of $T$ (b) at $\gamma_0=0.8$ [see also \ref{fig:idperf_LG_qbbat}(b,d,f)] and (c) at $\gamma_0= 2$. Results for the $s$- and $l$-JQB are shown for $\theta=0$ (maximum efficiency).}
\label{fig:effergo_qb}}
\end{figure}

{So far we have discussed the performance of the $p$-JQB by referring to the parametric plots in Fig. \ref{fig:idperf_LG_qbbat}, which are useful for identifying the optimal trade-off between efficiency and extracted work for each protocol but are less suited for comparing the different protocols under the same working conditions, $(\gamma_0,T)$, in which case $\mathcal{E}^{(g)} \geq \mathcal{E}^{(l)}$. For such a comparison, we introduce the product of the efficiency, $\eta^{(p)}(\theta)$, and the extracted work, $\mathcal{E}^{(p)}$, as a figure of merit encompassing both quantities in a single trade-off quantity. In terms of the latter---that we call \textit{efficient extracted work}, similarly to the efficient power in heat engines~\cite{Yilmaz16, Singh20}---we observe that the local extraction can outperform the global one for sufficiently large coupling strength, $\gamma_0$, and low temperature (Fig. \ref{fig:effergo_qb}).}

{Finally, we elaborate on admissible configurations of the specific superconducting setup discussed in Sec. \ref{sec:min_model_squid} in relation to our assumptions and to the results  discussed above.}
First, we provide a possible estimate of the timescale over which ergotropy is extracted from the single qubit in the $s$-JQB. The unitary \eqref{eq:Uerg_qiskit} which accomplishes this task is the product of two rotations $R_a(\phi) = e^{i \phi \sigma_a /2}$, with $a=x,y,z$, one about the $y$-axis (responsible for extracting ergotropy) and one about the $z$-axis (which does not affect the ergotropy). Therefore, assuming $\theta = 0$, we can ultimately focus on $R_{y}(2\alpha)$, which can be decomposed in terms of rotations only about the $x$- and $z$-axis, $R_{y}(2\alpha) = R_{z}(3\pi)R_{x}(\pi/2)R_{z}(\pi+2\alpha)R_{x}(\pi/2)$, each of which can be realized using finite duration voltage bias pulses~\cite{Griffith_2007}, with corresponding frequencies  $\nu_x = E_J/h$ and $\nu_z = E_C(n_g-1/2)/h$. Considering pulses of duration of the order $\sim 100$ ps, to account for possible experimental constraints on the shortest duration of a pulse \cite{Bladh_2005}, we can estimate the timescale of extraction of ergotropy as $\sim 1$ ns, much shorter than the typical longitudinal and transverse relaxation times of superconducting charge qubits $T_1 \sim 1 \div 10$ $\mu$s and $T_2\sim 0.1 \div 1$ $\mu$s  \cite{You2005}, respectively. Since thermalization involves relaxation processes, our assumption that the only role played by the reservoir is to thermalize the system at the end of every cycle is consistent with the time scales estimated above.
This assumption works when the temperature is sufficiently low. 
For the superconducting setup introduced in Sec. \ref{sec:squid}, this is the case when 
${k_B}T \ll E_C$, being a typical temperature
{$k_B T/h \sim 0.625\,\rm{GHz}$ \cite{Nakamura1999} and $E_C/h \sim 10 \div 100 \, \rm{GHz}$.}
In this regard, we want to stress that the results in Fig. \ref{fig:idperf_LG_qbbat}, and in the following, are expressed in units of $\Omega = E_{C}\left( n_{g}-1/2\right)$. Therefore, the fact that $\Omega$ depends also on the dimensionless gate charge $n_{g}$, whose value can be tuned through the gate voltage of the Cooper pair boxes \cite{Vion_2002}, allows for investigating regimes not limited to ${k_B}T \ll \Omega$, while keeping ${k_B}T \ll E_C$ (the low temperature regime).

{ A possible experimental scenario consistent with  Fig. \ref{fig:idperf_LG_qbbat}(a) at fixed $T = 0.5\Omega / k_B = 30$ mK \cite{Nakamura1999} is obtained for $C_J = 900$ aF, $C_g = 1.65$ aF, $V_g = 100$ mV, resulting in $n_g \approx 0.515$, $E_C/h \approx 86$ GHz and $\Omega/h = 1.28$ GHz. These values, for $E_J/h \sim 10$ GHz, allow to explore the range $\vert \gamma_0 \vert \leq 7.8$ while fulfilling $E_J,k_B T \ll E_C$. On the other hand, the same set of parameters provides $\gamma_0 \approx 0.78$ for $E_J/h \approx 1$ GHz, consistently with $\gamma_0=0.8$ in Fig. \ref{fig:idperf_LG_qbbat}(b), where $T \leq 10\Omega$ still fulfills $k_B T \ll E_C$.}


\subsection{Simulation on superconducting quantum computers}
\label{sec:jqb_sim}
After theoretically investigating the performance of the JQB in ideal conditions, we have simulated its functioning on a quantum computer.
In order to do so, we have used the quantum simulator \texttt{aer\_simulator}, without noise on the gates, and the quantum machines \texttt{ibm\_cairo}\footnote{This device was officially retired by IBM on April 30, 2024.} (with restrained access), which is one of the IBM Quantum Falcon processors with 27 qubits, and \texttt{ibm\_brisbane} (with open access),  which is one of the IBM Quantum Eagle processors with 127 qubits. All these platforms are accessible via the Qiskit package~\cite{qiskit2024}. 
{ These platforms, although based on superconducting qubits, differ from the one we propose in Sec. \ref{sec:min_model_squid} for our cyclic quantum battery. Therefore, it is worth commenting preliminarily about the different setups and outlining how the simulation is carried out.

The minimal model for a cyclic quantum battery in Fig. \ref{fig:dcsquids} is based on dc-SQUID charge qubits, designed to operate in the regime $E_J \ll E_C$. These charge qubits, however, are not suited for quantum computing, because environmental charge noise  strongly limits their coherence times \cite{Kjaergaard2020}. For this reason many superconducting quantum computing platforms, including IBM quantum hardwares, are currently based on transmons, an evolution of superconducting charge qubit which leverages the regime $E_J \gg E_C$, achieved by shunting the Josephson junction with an additional large capacitor~\cite{Koch_2007}, to improve coherence times by reducing the sensitivity to charge noise.} 
{ Although transmons and dc-SQUID charge qubits differ in their specific design, they are akin in their superconducting nature. Therefore, we opted for a transmon-based quantum hardware over other available quantum computing hardwares (e.g., photonic \cite{Slussarenko2019} or trapped-ion based \cite{Bruzewicz2019} ones) 
because we think it provides a more suitable and meaningful platform to simulate the proposed superconducting quantum battery. }

{Unless one has access to low-level controls on the hardware, the operations one can perform on a quantum computer are limited to unitaries and measurements. Considering this constraint and the different setup of IBM quantum processors with respect to ours, simulating the proposed thermodynamic cycle does not consist in rigorously performing each stroke, e.g, connection and disconnection, but rather in preparing the two-qubit states required to assess the energetics of the cycle. The simulation is thus carried out as a two-step process: (i) The initial Gibbs state is prepared in the quantum register and a quantum state tomography (QST) is performed; (ii) The initial Gibbs state is prepared again from scratch, it undergoes the unitary work extraction, and then a QST is performed. The QSTs are required to reconstruct the states that are actually implemented in the quantum register before and after work extraction. Such states, in turn, allow us to determine the proper unitary for work extraction and the energetics of the cycle, upon computing the expectation values of the Hamiltonian after each stroke as per Sec. \ref{sec:theo_td_cycle}.

In this framework, the parameters characterizing the quantum battery, i.e., the temperature $T$ and the coupling strength $\gamma_0$, are thus implicitly encoded in the preparation of the initial Gibbs state.}

\begin{figure*}[!htb]
    \centering
	\includegraphics[width=\columnwidth]{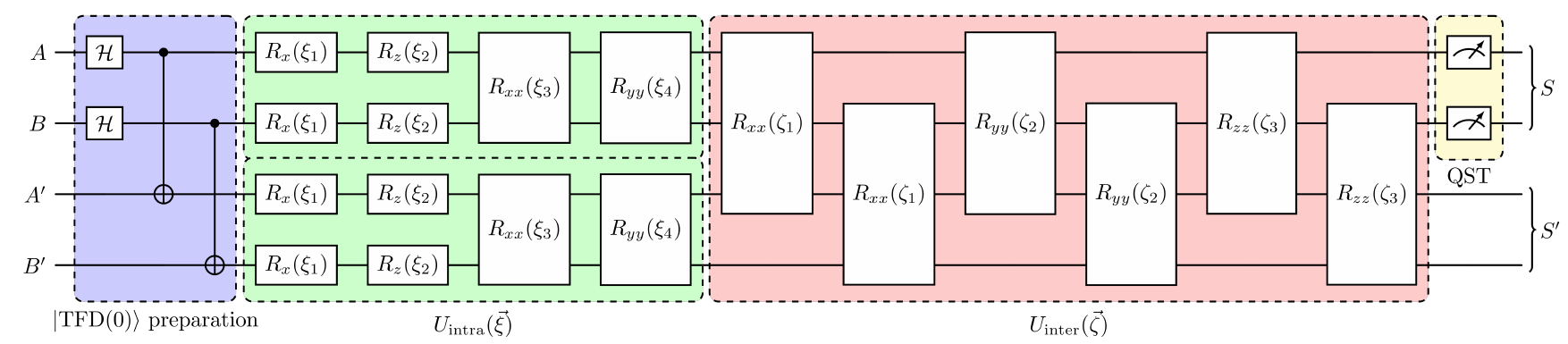}
	\caption{Quantum circuit associated to a single-step of the protocol that determines the optimal parameters to initialize the system $S$ in the Gibbs state. Blue area: preparation of the state $\ket{\rm{TFD}(0)}$ in Eq. \eqref{eq:TFD_0}; green area: unitary operation acting separately on the system $S$ and $S'$; red area: unitary operation coupling the system $S$ and $S'$; yellow area: quantum state tomography to determine the state of the system $S$. Here, we have used the gates $R^J_i(\theta)=e^{-i\frac{\theta}{2}\sigma_J^i}$ and $R^{JK}_{ii}(\theta)=e^{-i\frac{\theta}{2}(\sigma_J^i\otimes\sigma_K^i)}$ with $i=x,y,z$ and $J,K=A,A',B,B'$ to implement the unitaries introduced in the main text.}
	\label{fig:TFD_circuit}
\end{figure*}


\subsubsection{Generation of the initial Gibbs state}
The first, and most critical, step needed to simulate a thermodynamic cycle like the one discussed above, is the preparation of the initial Gibbs state $\tau_{\beta}$. Indeed, quantum computers are specifically designed to perform reversible unitary quantum operations and measurements~\cite{Benenti_book}. Therefore, it is not straightforward for an end user to realize mixed states using them. A possible way to overcome this issue is to rely on the so called thermofield double (TFD) states~\cite{Wu_prl2019, Zhu_pnas2020, Sagastizabal_npj2021, Consiglio24}. The idea behind this approach is to generate a pure state (the TFD state) in an enlarged system made up of two identical copies of the quantum system under investigation in such a way that tracing out the degrees of freedom of one of the two copies the other is described by the desired Gibbs state. In the following, we will use the notation introduced above for the system under investigation $S$ and we will add a prime index to distinguish the ancillary system $S'$. 

To proceed further in this direction, we consider the Hamiltonian $H$ in Eq. \eqref{eq:Ham_gamma}, defined in the Hilbert space of the system $S$, and a formally identical Hamiltonian $H'$ defined in the Hilbert space of the ancillary system $S'$. The TFD state is then defined on the enlarged Hilbert space and can be written as 
\begin{equation}
    \ket{\rm{TFD}(\beta)}=\sum_{k=1}^d \frac{e^{-\frac{\beta \varepsilon_k^\uparrow}{2}}}{\sqrt{\Tr[e^{-\beta H}]}}\vert \epsilon_k^\uparrow \rangle_S\vert \epsilon_k'^{\uparrow} \rangle_{S'}.
    \label{eq:TFD_state}
\end{equation}
As can be expected, by tracing out the degrees of freedom of the ancillary system $S'$, one recovers the Gibbs state $\tau_{\beta}$ for the system $S$ introduced in Eq. \eqref{eq:tau_gibbs}. Moreover, in the specific case under investigation we have $d=4$.

To realize the state in Eq. \eqref{eq:TFD_state} we first apply two Hadamard gates ($\mathcal{H}$) and two CNOT gates (blue region in the scheme of Fig. \ref{fig:TFD_circuit}) to initialize the enlarged system in the state 
\begin{eqnarray}
\ket{\rm{TFD}(0)} = &\frac{1}{2}\left(\ket{0}_A\ket{0}_{A'}+\ket{1}_A\ket{1}_{A'}\right)\nonumber\\
&\otimes \left(\ket{0}_B\ket{0}_{B'}+\ket{1}_B\ket{1}_{B'}\right)&. 
\label{eq:TFD_0}
\end{eqnarray}
We then use a variational scheme with an ansatz motivated by quantum-approximate optimization algorithms \cite{Blekos_2024} where intra-system and inter-system unitary operations are alternated. The former class of unitaries (green region in 
Fig. \ref{fig:TFD_circuit}) acts separately to $S$ and $S'$ and can be written as
\begin{eqnarray}
    U_{\rm{intra}}(\vec{\xi})
    & = e^{-i\frac{\xi_4}{2}(\sigma_A^{y}\otimes\sigma_B^{y}+\sigma_{A'}^{y}\otimes\sigma_{B'}^{y})}\nonumber\\
    & \times e^{-i\frac{\xi_3}{2}(\sigma_A^{x}\otimes\sigma_B^{x}+\sigma_{A'}^{x}\otimes\sigma_{B'}^{x})}\nonumber\\
    & \times e^{-i\frac{\xi_2}{2}(\sigma_A^{z}+\sigma_B^{z}+\sigma_{A'}^{z}+\sigma_{B'}^{z})}\nonumber\\
    & \times e^{-i\frac{\xi_1}{2}(\sigma_A^{x}+\sigma_B^{x}+\sigma_{A'}^{x}+\sigma_{B'}^{x})},
    \label{U_intra}
\end{eqnarray}
while the latter (red region in 
Fig. \ref{fig:TFD_circuit}) couples $S$ and $S'$ and reads
\begin{eqnarray}
    U_{\rm{inter}}(\vec{\zeta})
    & = e^{-i\frac{\zeta_3}{2}(\sigma_A^{z}\otimes\sigma_{A'}^{z}+\sigma_B^{z}\otimes\sigma_{B'}^{z})} \nonumber\\
    & \times  e^{-i\frac{\zeta_2}{2}(\sigma_A^{y}\otimes\sigma_{A'}^{y}+\sigma_B^{y}\otimes\sigma_{B'}^{y})}\nonumber\\
    & \times e^{-i\frac{\zeta_1}{2}(\sigma_A^{x}\otimes\sigma_{A'}^{x}+\sigma_B^{x}\otimes\sigma_{B'}^{x})}. 
    \label{U_inter}
\end{eqnarray}
Notice that the form of Eq. (\ref{U_intra}) is motivated by the single- and two-qubit terms appearing in the Hamiltonian in Eq. (\ref{eq:Ham_gamma}), while Eq. (\ref{U_inter}) is built extending the same idea to include coupling among the $S$ and $S'$~\cite{Zhu_pnas2020, Sagastizabal_npj2021}.

Altogether they lead to the unitary 
\begin{equation}
    U_{\rm TFD}(\vec{\xi},\vec{\zeta})=U_{\rm{inter}}(\vec{\zeta})U_{\rm{intra}}(\vec{\xi}),
\end{equation}
where the variational parameters $\vec{\xi}=(\xi_1,\xi_2,\xi_3,\xi_4)$ and $\vec{\zeta}=(\zeta_1,\zeta_2,\zeta_3)$ need to be optimized to generate the closest state to the target state $\ket{\rm{TFD}(\beta)}$ according to a given cost function. 
{The variational parameters implicitly encode the temperature $T$ and the coupling strength $\gamma_0$ of the quantum battery.} In the present case, we have chosen as cost function the infidelity with respect to the target Gibbs state defined as $1-F(\tilde{\tau}(\vec{\xi},\vec{\zeta}),\tau_{\beta})$
 with 
\begin{equation}
    \tilde{\tau}(\vec{\xi},\vec{\zeta})= \Tr_{S'}[U_{\rm{TFD}}(\vec{\xi},\vec{\zeta})\ket{\rm{TFD}(0)}\bra{\rm{TFD}(0)}U^{\dag}_{\rm{TFD}}(\vec{\xi},\vec{\zeta})]
\end{equation}
and where
\begin{equation}
    F(\rho,\sigma)=\Tr[\sqrt{\sqrt{\rho}\sigma\sqrt{\rho}}]
\end{equation}
is the conventional definition of the fidelity for two arbitrary density matrices $\rho$ and $\sigma$~\cite{Nielsen_2010book}. For completeness, we have also tested other cost functions such as the free energy and weighted sums of the infidelity with various normalized energy components (not shown). However, the latter approaches were not as good in estimating the extracted work and the efficiency of the considered system as the approach based only on the infidelity.

To evaluate the fidelity, we perform a quantum state tomography (QST) (yellow region in the scheme of Fig.~\ref{fig:TFD_circuit})~\cite{Nielsen_2010book}. This is an experimental procedure that allows to reconstruct the density matrix of a quantum state. It consists in preparing the system in the same state many times and measuring it in a tomographically complete basis of measurement operators. That is, the measured operators must form an operator basis on the Hilbert space of the system. 

\begin{figure}[!tb]
    \centering
    \includegraphics[width=0.7\columnwidth]{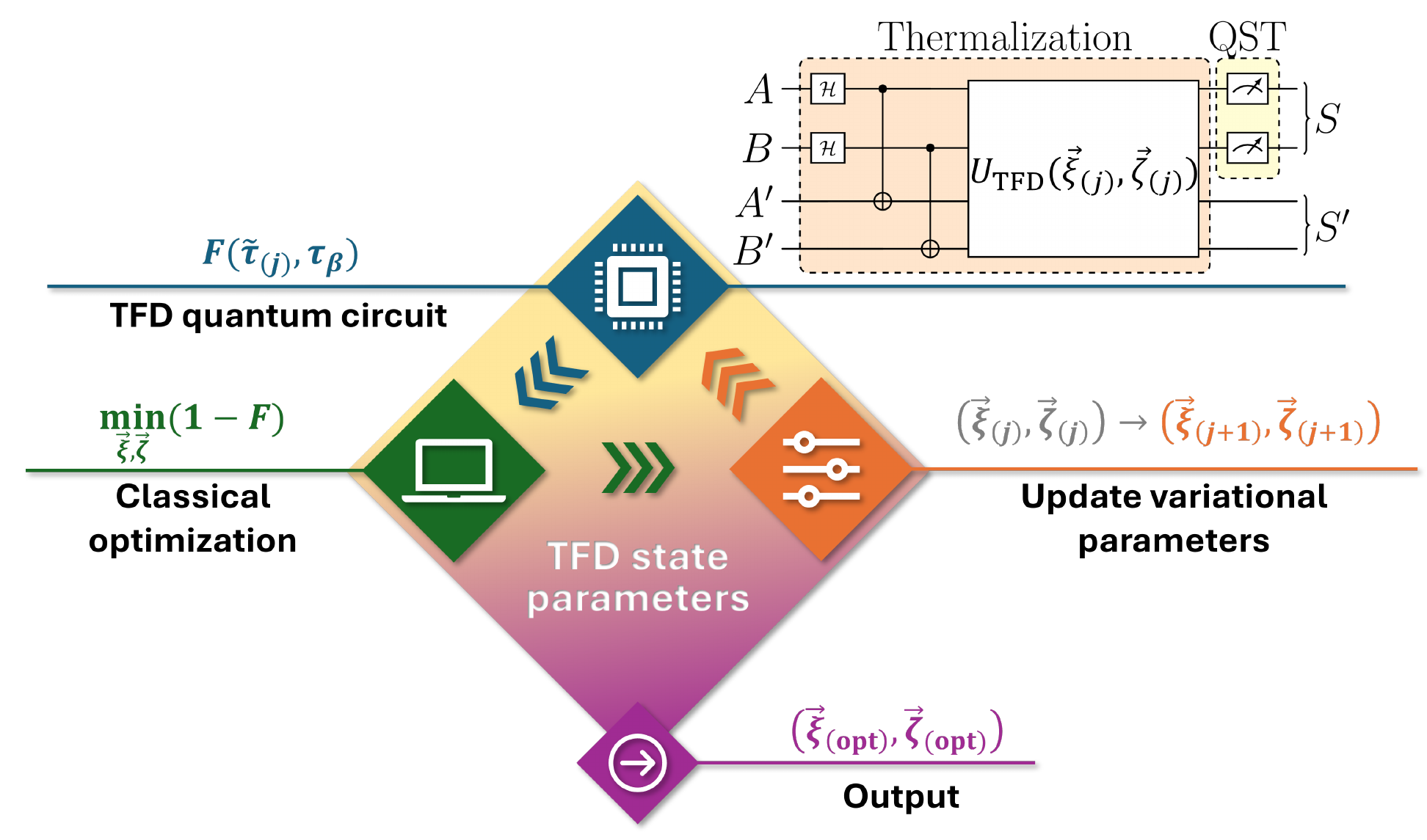}
    \caption{Scheme of the variational algorithm used to prepare the TFD state. In each iteration $(j)$ the quantum circuit in Fig. \ref{fig:TFD_circuit} is executed to determine the fidelity $F$ between the state $\tilde{\tau}_{(j)}\equiv \tilde{\tau}(\vec{\xi}_{(j)},\vec{\zeta}_{(j)})$ and the target Gibbs state, $\tau_{\beta}$ (blue part). The infidelity ($1-F$) is then minimized via a Bayesian optimization, performed on a classical computer (green part), in order to update the circuit parameters in $(\vec{\xi}_{(j+1)},\vec{\zeta}_{(j+1)})$ (orange part) and start a new iteration, $(j+1)$. This procedure is iterated until the algorithm converges,  yielding the optimal parameters $(\vec{\xi}_{\rm (opt)},\vec{\zeta}_{\rm (opt)})$ (violet part).
    }
    \label{fig:QOAOscheme}
\end{figure}

To obtain the best estimation of the variational parameters $\vec{\xi}$ and $\vec{\zeta}$, the steps discussed above and illustrated in Fig.~\ref{fig:TFD_circuit} were implemented using the simulator without noise with number of shots equal to $1000$.
After every QST the optimization of the parameters was performed via a Bayesian optimisation~\cite{scikit-optimize, garnett_2023book}, an approach which is particularly convenient when the cost function to optimize is a black box for which no closed form is known (nor its gradients), is computationally expensive to evaluate and noisy. These conditions perfectly fit the case under investigation. The achieved parameters were then fed back into the quantum circuit of Fig.~\ref{fig:TFD_circuit} starting another round of optimization. This procedure was iterated for a number of times sufficient to reach a convergence of the results for the density matrix of $S$ (typically $400$-$600$). In Fig. \ref{fig:QOAOscheme} we report a scheme of the whole optimization algorithm.

\subsubsection{Simulation of the thermodynamic cycle}
Once the parameters leading to the best approximation of the thermal state for $S$ were obtained, we moved on to the simulation of the whole thermodynamic cycle, both using the simulator and the two real devices mentioned above. One purpose of this simulation is to perform a feasibility test of the protocols based on the extraction of local ergotropries. Since the principle is equivalent for both the $s$- and the $l$-JQB---ergotropy is locally extracted from each qubit by single-qubit gates, which can be eventually executed in parallel---, without loss of generality we restrict the discussion to the $s$-JQB. Note that during strokes \textsc{(i$\rightarrow$ii)} and \textsc{(iii$\rightarrow$iv)} the density matrix of the bipartite system does not change. This means that we do not have to perform any operation on the qubits to simulate these strokes.

\begin{figure}[!tb]
\centering
\includegraphics[width=0.7\columnwidth]{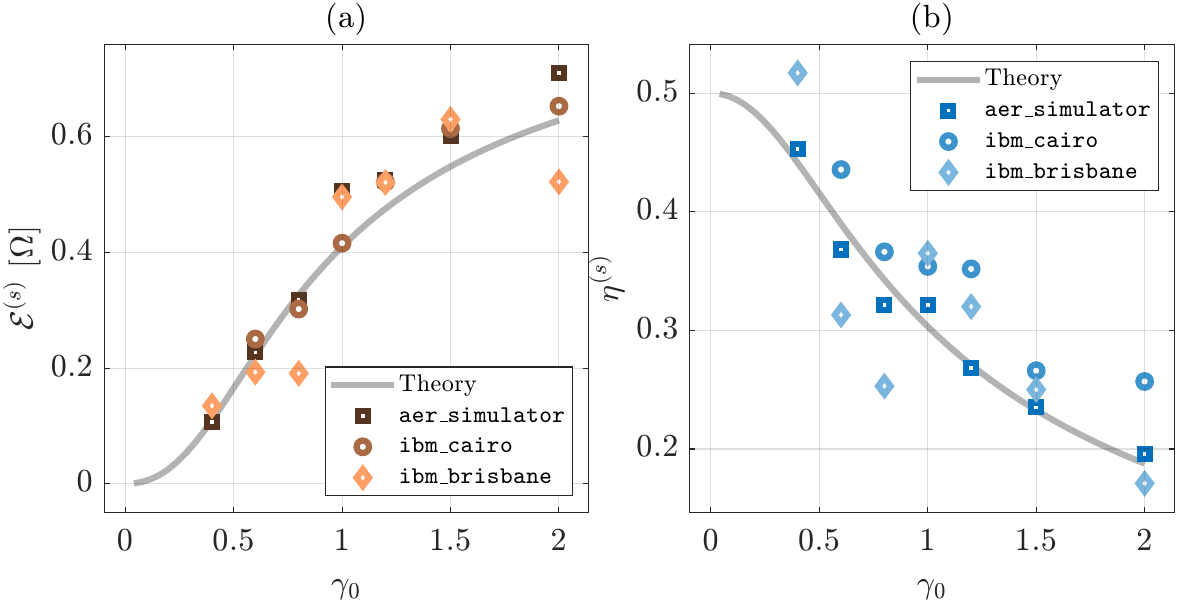}
\caption{Performance of the simulated $s$-JQB at $T = 0.5\Omega$ and $\theta = 0$. (a) Extracted work $\mathcal{E}^{(s)}=\mathcal{E}$ and (b) efficiency $\eta^{(s)}$ as a function of $0<\gamma_0 \leq 2$ for the ideal cycle starting with a Gibbs state (gray curve) and for the cycles starting with a thermofield double state (markers) simulated on (\texttt{aer\_simulator}) and actual quantum hardwares (\texttt{ibm\_cairo} and \texttt{ibm\_brisbane}).
\label{fig:exp_1qbbat}}
\end{figure}

\begin{figure}[!tb]
    \centering
    \includegraphics[width=0.5\columnwidth]{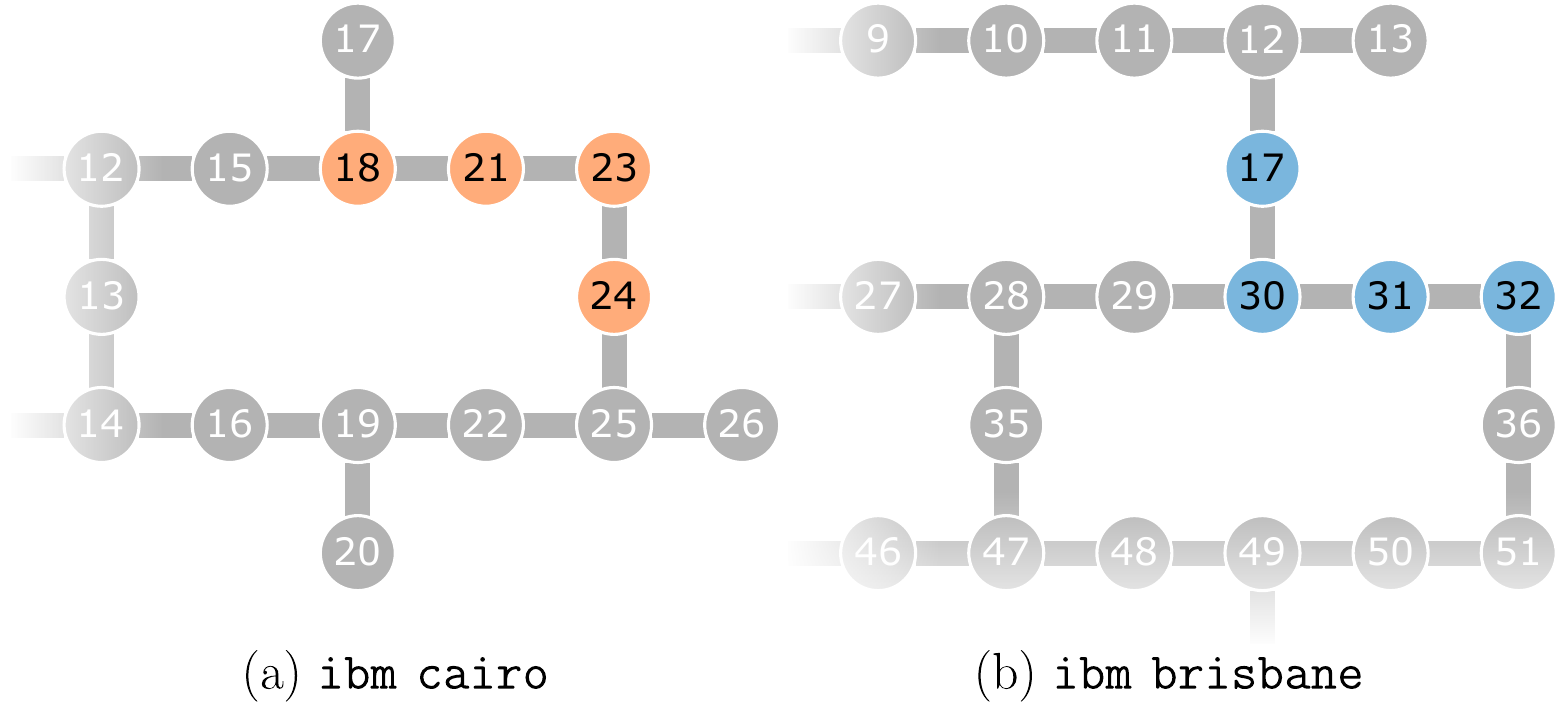}
    \caption{Qubit connectivity map of (a) \texttt{ibm\_cairo} and (b) \texttt{ibm\_brisbane}, with the qubits used for the simulations highlighted in orange and blue, respectively.} 
    \label{fig:ibm_cairo_brisbane}
\end{figure}

Here, the extraction of ergotropy from the single qubit is straightforwardly implemented on the quantum computer as the product of two single-qubit rotation gates, see Eq.~\eqref{eq:Uerg_qiskit}. For the simulation, we consider the regime where the best performance is ideally expected, i.e., $\gamma_0 >0$ with $\theta=0$ [Fig.~\ref{fig:idperf_LG_qbbat}(a,b)]. The latter assumption further simplifies the unitary \eqref{eq:Uerg_qiskit} to a single-qubit rotation about the $y$-axis. In this regard, the parameter $\alpha$ was directly evaluated from the coordinates on the Bloch sphere of the density matrix achieved after the thermalization.
Figure~\ref{fig:exp_1qbbat} shows the performance of the simulated $s$-JQB in terms of the extracted work $\mathcal{E}^{(s)}$ [panel (a)] and of the efficiency $\eta^{(s)}$ [panel (b)] at a given temperature ($T = 0.5\Omega$) and for different values of the coupling $\gamma_0\in[0.4,2]$. We focused our analysis on this region of parameters to achieve a good signal-to-noise ratio, as indicated by the ideal results shown in Fig.~\ref{fig:idperf_LG_qbbat}(a,b). For $\gamma_0 < 0.4$, we expect very small values of ergotropy, whereas for $\gamma_0 > 2$, we expect low cycle efficiencies, making the measurement of these quantities more challenging. Moreover, for $T > \Omega$, we observed issues with the convergence of the variational algorithm when preparing the initial Gibbs state. Conversely, for $T < \Omega$ we do not expect a strong dependence of the ergotropy and the efficiency on $T$, therefore, we have fixed $T=0.5\Omega$. 
{Despite these limitations in the choice of viable parameters for the simulation, we point out that those we considered are consistent with the experimentally admissible configuration of the dc-SQUID setup discussed at the end of Sec. \ref{sec:id_perf} in relation to Fig. \ref{fig:idperf_LG_qbbat}(a).} The simulator data were generated by using 1000 shots and averaging over 30 runs. In contrast, the data from the real machine were obtained from a single run with 4000 shots. Note that reproducing the circuit in Fig.~\ref{fig:TFD_circuit} requires 4 qubits with a square connection, which is not available in the devices we have used. To minimize the necessary SWAP gates, we utilized 4 qubits connected in a linear configuration (the mapping of qubits $[A,B,A',B']$ is $[18,24,21,23]$ for \texttt{ibm\_cairo} and $[32,17,31,30]$ for \texttt{ibm\_brisbane}, see Fig. \ref{fig:ibm_cairo_brisbane}).

For both the ergotropy and the efficiency the agreement between the data obtained from the simulator and the two considered real quantum devices and the theoretical curves is good. We observe, however, that the agreement for the ergotropy [panel (a)] is better than that for efficiency [panel (b)]. The reason of this can be traced back to the definitions of the considered quantities. Indeed, after performing the QST to obtain the density matrix of the system $S$ before and after the ergotropy extraction, accomplished by applying the rotation in Eq.~\eqref{eq:Uerg_qiskit}, the extracted work $\mathcal{E}^{(s)}$ is numerically evaluated in a unique step. Instead, assessing the efficiency $\eta^{(s)}$ requires multiple steps. First, it requires the separate calculation of $\mathcal{E}^{(s)}$, $E_c^{(s)}$, and $E_d$. These quantities are then combined to construct $\eta^{(s)}$ as in Eq. \eqref{eq:eta}. When performing calculations with quantities affected by errors, these errors propagate through each step, consequently, the individual errors from $\mathcal{E}^{(s)}$, $E_c^{(s)}$, and $E_d$ combine and potentially accumulate, leading to a greater overall error in the final calculation of $\eta^{(s)}$.

In this regard, errors have different possible origins. One source of error is associated to the parameter optimization required to simulate the initial Gibbs state $\tau_{\beta}$. In the cases shown here, the fidelity between the ideal Gibbs state and the one obtained following the optimization procedure ranges between $0.95$ and $0.98$. This error affects all data at a given $\gamma_0$ in the same way, since the parameter optimization is performed once for each value of $\gamma_0$ using the \texttt{aer\_simulator}. The other main sources of error are: (i) the intrinsic noise of the quantum computers (gates, readout, and qubit decoherence) and (ii) the probabilistic nature of the measurements, performed in a finite number. For the considered IBM quantum devices the qubit state stability is of the order of $100 \, \mu\rm{s}$, the gates acting on the qubits have median single-qubit gate error $\sim 10^{-4}$, while the median two-qubit gate error and the median readout error are $\sim 10^{-2}$~\cite{qiskit2024}. As regard the probabilistic nature of the measurements we recall that we have performed $1000$ shots repeated for $30$ runs on the simulator and $4000$ shots for a single run on the real devices.

As it clearly emerges from this brief description, the sources of error affecting the simulation are numerous and of various nature. For this reason, estimating the error affecting the data reported in Fig. ~\ref{fig:exp_1qbbat} starting from these sources is very challenging. We can, however, estimate the magnitude of such errors starting from another observation. As already mentioned, the fidelity between the ideal Gibbs state and the one obtained on the simulator ranges from $0.95$ to $0.98$ (average $0.97$). Similarly, we have that the fidelity between the ideal Gibbs state and the one obtained on real machines ranges from $0.85$ to $0.97$ (average $0.91$). In the first case, the deviation from $1$ of the fidelity is attributed to the error in determining the parameters to construct the initial state, while in the second case all the errors described above combine. At this point, it seems reasonable to assume the parameter $1-F$, with $F$ the fidelity, as the relative error on each QST, thus on the determination of each density matrix. This will therefore be of the order of $3\%$ for the simulators and of $10\%$ for the real devices. Note that the argumentation has been made on the state before the extraction of ergotropy, but it is legitimate to extend it also to the state after the extraction of ergotropy since the difference associated to the $\sigma_y$ rotation (corresponding to 4 native single-qubit gates) on the entire process can be safely neglected. Propagating the errors through the usual rules, we can estimate a relative error of about $20\%$ for the ergotropy and of about $40\%$ for the efficiency. This estimation appears reasonable when compared with the obtained results (see Fig.~\ref{fig:exp_1qbbat}).

Finally, it is possible to estimate the extraction power associated to the considered quantum machines. The time required to perform the $\sigma_y$ rotation, associated with the extraction of ergotropy, is on the order of $100$ ns. In these devices, the qubits have a typical frequency of $5$ GHz, which corresponds to a typical energy level spacing
$\sim 20.6 \,\mu\mathrm{eV}$.
In passing, we notice that this value is of the same order of magnitude of what estimated for the JQB in Section~{\ref{sec:rjqb}. From Fig. \ref{fig:exp_1qbbat}, it emerges that the extracted energy spans in the range $0.2\div 0.6\,\,\Omega$, with $2\Omega$ being the energy level spacing of the qubit. Combining these data, it follows that the extracted power is on the order of $10^{-2} \,\rm{fW}$. As a final remark, we point out that the typical longitudinal, $T_1$, and transverse, $T_2$, relaxation times of the transmon superconducting qubits of the considered quantum machines are of the order $\sim 100$ $\mu$s, much longer than the time it takes to extract ergotropy  ($\sim 100$ ns).


\section{Conclusion}
\label{Conclusions}
We studied the concept of a cyclic quantum battery operating according to a given thermodynamic cycle. During the whole cycle, the setup is weakly interacting with a thermal bath, while the operations are carried out on timescales shorter than those of the thermalization. This guarantees that the only role played by the reservoir is to reinitialize the system to the same thermal state at the end of every cycle. Such state is no longer passive after switching off the interaction 
internal to the battery and therefore ergotropy extraction is possible.

Our proposal for a quantum battery is based on two connected qubits, whose non-trivial correlations, together with phase coherence, can be exploited to improve the performance of the device. Indeed, it is possible to apply local unitary operations on one or two qubits separately or acting globally on them to get useful energy. Hence, a key outcome of our work is that the efficiency of the setup in the case of parallel energy extraction from two qubits can exceed that where global operations are performed. Since local unitary operations are typically easier to implement than global ones, this result provides fertile ground for future implementations of such a kind of quantum battery.

To support future experimental realizations of the discussed quantum battery, we provided a minimal feasible scheme for a superconducting quantum battery based on coupled charged qubits which can be recharged following the proposed thermodynamic cycle. The major advantage of experimental schemes of this type, based on Josephson junctions, is their scalability, which leads to an enhancement of the overall storage capacity of the device. One could further develop this work by analyzing the impact of many-body correlations emerging by increasing the circuit size. 

Additionally, and this is a significant outcome of the present work, we simulated the proposed thermodynamic cycle on an IBM quantum hardware. The simulation on a quantum computer required to overcome several technical challenges, such as the realization of a given initial thermal state for the quantum battery and error mitigation. While the former was tackled with the introduction of thermofield double states, a Bayesian optimization procedure was needed to deal with the latter. Despite unavoidable errors, mainly caused by the intrinsic noise in the device and by the probabilistic nature of the measurements, the obtained results foster the implementation of a cyclic quantum battery using solid-state platforms. 

The reason why superconducting quantum hardware appears to provide a very suitable ground for developing our cyclic quantum batteries scheme is two-fold. First, the coupling to the environment is weak, in that the qubit longitudinal and transverse relaxation times are much longer than the ergotropy extraction time; next, entangling quantum gates could be used to exploit the quantum advantage in the charging time of quantum batteries. The present work therefore opens up new perspectives in the field of quantum batteries and more generally provides a wider spectrum of possibilities for the study of energy and information manipulation in superconducting quantum circuits.

\ack D.F. acknowledges the contribution of the European Union-NextGenerationEU through the ``Quantum Busses for Coherent Energy Transfer'' (QUBERT) project, in the framework of the Curiosity Driven 2021 initiative of the University of Genova. G.B, D.F., I.K., and L.R acknowledge support from the project PRIN 2022 - 2022XK5CPX (PE3) SoS-QuBa - ``Solid State Quantum Batteries: Characterization and Optimization'' funded within the programme ``PNRR Missione 4 - Componente 2 - Investimento 1.1 Fondo per il Programma Nazionale di Ricerca e Progetti di Rilevante Interesse Nazionale (PRIN)'', funded by the European Union - Next Generation EU. L.R.. I.K., and G.B. acknowledge support from the  Julian Schwinger Foundation (Grant JSF-21-04-0001) and from INFN through the project “QUANTUM”.

{We thank the two anonymous reviewers whose comments/suggestions helped to improve and clarify this manuscript.}

We acknowledge the use of IBM Quantum services and the access to them granted by INFN for this work. The views expressed are those of the authors, and do not reflect the official policy or position of IBM or the IBM Quantum team.


\appendix
\section{Pauli matrices and qubit states}
\label{app:eig_rho_qubit}
In this Appendix we recall some basics of the mathematical description of a qubit.
To begin, we provide the conventional expressions for the Pauli matrices
\begin{equation}
\sigma_x =
\left(
\begin{array}{cc}
0 & 1\\
1 & 0
\end{array}\right),\quad
\sigma_y =
\left(
\begin{array}{cc}
0 & -i\\
i & 0
\end{array}\right),\quad
\sigma_z =
\left(
\begin{array}{cc}
1 & 0\\
0 & -1
\end{array}\right),
\end{equation}
which are Hermitian, involutory, and unitary operators ($\sigma_a = \sigma_a^\dagger = \sigma_a^{-1}$, with $a=x,y,z$). Their commutation, $[\sigma_a,\sigma_b] = 2 i \varepsilon_{abc}\sigma_c$, and anticommutation relations, $\{\sigma_a,\sigma_b\} = 2 \delta_{ab} I$ with $I$ being the identity in $\mathbb{C}^{2\times 2}$, combine in
\begin{equation}
\sigma_a \sigma_b = \delta_{ab} I + i \varepsilon_{abc}\sigma_c.
\label{eq:pauli_comm_anti}
\end{equation}
Letting $\vec{\sigma} \equiv (\sigma_x,\sigma_y,\sigma_z)$, the following property applies,
\begin{equation}
e^{i\phi \hat{n} \cdot \vec{\sigma}} = I \cos\phi + i (\hat{n}\cdot \vec{\sigma}) \sin\phi 
\label{eq:exp_sigma_vec}
\end{equation}
with $\hat{n} \in \mathbb{R}^3$, $\vert \hat{n} \vert = 1$, and $\phi \in \mathbb{R}$.

The more general state of a qubit is represented by the density matrix \cite{Benenti_book}
\begin{equation}
\rho = \frac{1}{2}(I+ \vec{r} \cdot \vec{\sigma}) =  \frac{1}{2}
\left(
\begin{array}{cc}
1+z		& x - i y\\
x + i y	& 1-z
\end{array}\right),
\end{equation}
with $\vec{r}=(x,y,z) \in \mathbb{R}^3$ where
\begin{equation}
x = \Tr[\sigma_x \rho], \quad
y = \Tr[\sigma_y \rho], \quad
z = \Tr[\sigma_z \rho]
\label{eq:bloch_coords}
\end{equation}
are the coordinates of the state $\rho$ in the Bloch sphere. The radius $r \equiv \vert \vec{r}\vert = \sqrt{x^2+y^2+z^2}$ is related to the purity of the state $\Tr[\rho^2]=\frac{1}{2}(1+r^2)$, ranging from $r=0$ (maximally mixed state) to $r=1$ (pure state). Note that, for a real state $\rho$, one has $y=0$. The eigenvalues of $\rho$ are $\lambda_\pm  = \frac{1}{2}(1 \pm r)$ and the corresponding eigenstates are
\begin{equation}
\vert \lambda_\pm \rangle  = \mathcal{N}_{\pm}\left( \frac{x - i y}{\pm r - z} \vert 0 \rangle + \vert 1 \rangle \right),
\label{eq:eigvec_rho_def}
\end{equation}
with $\mathcal{N}_{\pm}$ being a proper normalization constant. Since $\lambda_- \leq \lambda_+$, as $0 \leq r \leq 1$, the eigenstate $\vert \lambda_- \rangle$ is the least populated, and $\vert \lambda_+ \rangle$ the most populated.


\section{Proof of Eq. \eqref{eq:Uerg_qiskit}}
\label{app:U_erg}
In this Appendix we prove that in the case the initial state of a qubit is described by a real density matrix, the unitary operator for ergotropy extraction $U(\theta)$ has the form given in Eq. \eqref{eq:Uerg_qiskit}. The proof below is based on~\cite{Barra_njp}.

For ease of notation, we omit the subscripts and denote by $H = \sigma_z$ and $\rho$ the local Hamiltonian $H_A$ and the reduced state of the qubit $\rho_{A,{\rm II}} = \Tr_B[\tau_\beta]$, respectively (recall $\varrho_{\rm II}=\varrho_{\rm I}=\tau_{\beta}$). In the unitary \eqref{eq:U_ergextr} a global phase can be factored out, and hence for a $(d=2)$-level system, there is only one independent phase. The latter, say the relative phase $0 \leq (\theta_A - \theta_B) \equiv 2\theta < 2 \pi$, is conveniently included in the unitary as follows
\begin{equation}
U(\theta) = e^{-i\theta} \vert \epsilon_1^\uparrow \rangle\langle \lambda_1^\downarrow \vert + e^{i\theta} \vert \epsilon_2^\uparrow \rangle\langle \lambda_2^\downarrow \vert,
\label{eq:Uerg_halfphase}
\end{equation}
with $0 \leq \theta < \pi$. Recalling the convention on the sorting of the eigenvalues, the eigensystem of $H$ is $\{ \vert \epsilon_1^\uparrow = -1 \rangle = \vert 1 \rangle, \, \vert \epsilon_2^\uparrow = +1 \rangle = \vert 0 \rangle\}$, and that of $\rho$ is $\{ \vert \lambda_1^\downarrow =\frac{1}{2}(1+r)\rangle = \vert \lambda_+ \rangle,\, \vert \lambda_2^\downarrow =\frac{1}{2}(1-r)\rangle = \vert \lambda_- \rangle\}$ with $0 \leq r \leq 1$ (see \ref{app:eig_rho_qubit}). 
{Note that the energy level spacing of the single qubit is $\epsilon_2^\uparrow - \epsilon_1^\uparrow = 2\Omega$, upon restoring $\Omega>0$ as the energy unit.}

Now we focus on the eigensystem of $\rho$. The density matrix $\rho$ is obtained as the reduced state of a Gibbs state. A Gibbs state \eqref{eq:tau_gibbs} is by definition a real symmetric matrix. As a result, reduced states obtained from a Gibbs state are in turn real and symmetric. Therefore, $\rho$ is a $2\times2$ real symmetric matrix, and the eigenvectors of such a matrix can be expressed in the form $\alpha \vert 0 \rangle + \beta \vert 1 \rangle$ and $-\beta \vert 0 \rangle + \alpha \vert 1 \rangle$, with $\alpha,\beta \in \mathbb{R}$. Accordingly, we can rewrite the eigenstates \eqref{eq:eigvec_rho_def} as
\begin{eqnarray}
\vert \lambda_2^\downarrow \rangle \equiv \vert \lambda_- \rangle & = \cos\alpha \vert 0 \rangle + \sin\alpha \vert 1 \rangle,\nonumber\\
\vert \lambda_1^\downarrow \rangle \equiv \vert \lambda_+ \rangle & = -\sin\alpha \vert 0 \rangle + \cos\alpha \vert 1 \rangle,
\end{eqnarray}
with
\begin{equation}
\cos\alpha = - \frac{x}{\sqrt{2r(r+z)}}, \quad \sin\alpha = \frac{r+z}{\sqrt{2r(r+z)}},
\end{equation}
where $r = \sqrt{x^2 + z^2}$ (recall that $y=0$, being $\rho$ real), from which $\tan\alpha = -(r+z)/x$. Note that $r+z \geq 0$ always. Then, we can write the unitary operator \eqref{eq:Uerg_halfphase} as
\begin{eqnarray}
U(\theta)
&= e^{i\theta} \cos\alpha \vert 0 \rangle\langle 0 \vert + e^{i\theta} \sin\alpha \vert 0 \rangle\langle 1 \vert\nonumber\\
&\quad -e^{-i\theta} \sin\alpha \vert 1 \rangle\langle 0 \vert + e^{-i\theta} \cos\alpha \vert 1 \rangle\langle 1 \vert ,
\end{eqnarray}
from which the matrix representation in the first equality of Eq. \eqref{eq:Uerg_qiskit} follows. Using the property \eqref{eq:exp_sigma_vec}, it is possible to rewrite it in terms of 
\begin{equation}
e^{i \theta \sigma_z} =
\left(
\begin{array}{cc}
e^{i \theta}& 0\\
0			&e^{-i \theta}
\end{array}\right),\quad
e^{i\alpha \sigma_y}  =
\left(
\begin{array}{cc}
 \cos\alpha	& \sin\alpha\\
-\sin\alpha	& \cos\alpha
\end{array}\right),
\end{equation}
from which the second equality of Eq. \eqref{eq:Uerg_qiskit} can be easily verified, concluding the proof.


\section{Details on the energetics of the working cycle of the JQB}
\label{app:ideal_energies}
In this Appendix, we provide the analytical derivation of the results presented in Sec. \ref{sec:energetics_wc_JQB}, regarding the energetics of the $p$-JQB for the different Protocols \ref{prot:s}--\ref{prot:g}, $p=s,l,g$, respectively. 
First, we focus on the energetics of the protocols $p=s,l$, in which ergotropy is locally extracted from qubits. We then address the energetics of the protocol $p=g$, which requires an extended discussion due to the presence of a degenerate energy level in the bare Hamiltonian $H_0=H_A+H_B$, from which the global ergotropy is extracted.

In the following, we denote by $\langle O \rangle \equiv \Tr[O \varrho_{\rm I}]$ the expectation value of the generic observable $O$ on the initial Gibbs state $\varrho_{\rm I}=\tau_\beta$ with Hamiltonian $H = H_0 + \gamma_0 H_{\rm int}$, and, for conciseness, we omit identities and the $\otimes$ symbol when there is no ambiguity from the context. We observe that $\langle \sigma_A^a \rangle = \langle \sigma_B^a \rangle$ and $\langle \sigma_A^a \sigma_B^b \rangle = \langle \sigma_A^b \sigma_B^a \rangle$ with $a,b = x,y,z$, since the Hamiltonian \eqref{eq:Ham_gamma}, hence the Gibbs state, is symmetric under exchange of the two qubits. Accordingly, the results that follow will be expressed in terms of $\langle \sigma_A^a \rangle$ and $\langle \sigma_A^a \sigma_B^b \rangle$.

Note that the disconnection energy \eqref{eq:ideal_Ed} is simply computed from definition \eqref{eq:Ed}, and, as such, remains the same for all the considered protocols. Hence, it will not be addressed in the following Sections.


\subsection[Energetics of the (p=s,l)-JQB]{Energetics of the $(p=s,l)$-JQB}
The ergotropy $\mathcal{E}$ \eqref{eq:ideal_Erg} extracted from the single qubit is given by \eqref{eq:theo_ergo}, recalling that the passive state of a qubit with respect to $\sigma_z$ is $\pi = \frac{1}{2}(1-r \sigma_z)$, and $x,z,r$ are defined in Eq. \eqref{eq:bloch_coords}. Hence, $\mathcal{E}^{(s)}=\mathcal{E}$ and $\mathcal{E}^{(l)}=2\mathcal{E}$ \eqref{eq:ErgLoc_theo} trivially follows.

The connection energy \eqref{eq:Ec} for the $s$-JQB reads (identifying $U_A(\theta_A)\equiv U(\theta)$ for notational convenience) 
\begin{eqnarray}
E_c^{(s)}(\theta) = &-\gamma_0\big(
\langle U^\dagger(\theta) \sigma_A^x U(\theta)\rangle
+\langle \sigma_B^x \rangle\nonumber\\
&+ \langle U^\dagger(\theta) \sigma_A^x U(\theta) \otimes \sigma_B^x\rangle
- \langle U^\dagger(\theta) \sigma_A^y U(\theta)\otimes \sigma_B^y\rangle
\big),
\end{eqnarray}
where the cyclic property of the trace has been used. Therefore, we have to compute $U^\dagger(\theta) \sigma_A^a U(\theta)$, with $a=x,y$ and $U(\theta)$ in Eq. \eqref{eq:Uerg_qiskit}. Using Eq. \eqref{eq:exp_sigma_vec}, and the property \eqref{eq:pauli_comm_anti}, by direct inspection it is possible to prove that
\begin{eqnarray}
U^\dagger(\theta) \sigma_A^x U(\theta) &=
\cos(2\theta)\cos(2\alpha)\sigma_A^x + \sin(2\theta)\sigma_A^y \nonumber\\
&\quad - \cos(2\theta)\sin(2\alpha)\sigma_A^z,\\
U^\dagger(\theta) \sigma_A^y U(\theta) &=
- \sin(2\theta)\cos(2\alpha)\sigma_A^x + \cos(2\theta)\sigma_A^y\nonumber\\
&\quad + \sin(2\theta)\sin(2\alpha)\sigma_A^z ,\\
U^\dagger(\theta) \sigma_A^z U(\theta) &=
\sin(2\alpha)\sigma_A^x + \cos(2\alpha)\sigma_A^z ,
\end{eqnarray}
where the latter is reported for completeness.
For a real state $\varrho_{\rm I}$ (as a Gibbs state is), $\langle \sigma_A^y\rangle = \langle \sigma_A^y \sigma_B^x\rangle
= \langle \sigma_A^y \sigma_B^z\rangle = 0$. Combining all previous results, Eq. \eqref{eq:ideal_Ec} follows.

Analogously, we can compute the connection energy for the $l$-JQB, recalling that the two-qubit unitary is $\mathcal{U}_l(\theta_A,\theta_B) = U_A(\theta_A)\otimes U_B(\theta_B)$  with $U_k(\theta_k)$  in Eq. \eqref{eq:Uerg_qiskit}, 
\begin{eqnarray}
E_c^{(l)}(\theta_A,\theta_B) = &-\gamma_0 \Big\{ \Big(\!\cos(2\theta_A)\!+\!\cos(2\theta_B)\!\Big) \Big(\! \cos(2\alpha) \langle\sigma_A^x\rangle \nonumber\\
& - \sin(2\alpha)\langle\sigma_A^z\rangle \!\Big)
+\cos{[2(\theta_A+\theta_B)]} \Big(\! \cos^2(2\alpha) \langle\sigma_A^x\sigma_B^x\rangle \nonumber\\
&-\langle\sigma_A^y\sigma_B^y\rangle +\sin^2(2\alpha)\langle\sigma_A^z\sigma_B^z\rangle  - \sin(4\alpha) \langle \sigma_A^x \sigma_B^z\rangle \!\Big)\Big\}.
\end{eqnarray}
Note that, in principle, such a unitary depends on two independent arbitrary phases. Now, we observe that
$\cos(2\alpha) \langle\sigma_A^x\rangle - \sin(2\alpha)\langle\sigma_A^z\rangle = 0$,
which can be easily proved by using Eq. \eqref{eq:bloch_coords} and, provided that $\alpha  \neq \frac{\pi}{2} + k \pi$ with $k \in \mathbb{Z}$,
\begin{eqnarray}
\cos(2\alpha ) &= \frac{1\!-\!\tan^2\alpha}{1+\tan^2\alpha}=-\frac{z}{r},\nonumber\\
\sin(2\alpha ) &= \frac{2\tan\alpha}{1+\tan^2\alpha}=-\frac{x}{r},
\label{eq:cos_sin_xzr}
\end{eqnarray}
because $\alpha = \arctan[-(r+z)/x]$ with $y=0$. Therefore, $E_c^{(l)}(\theta_A,\theta_B)=E_c^{(l)}(\theta_A+\theta_B)$ is a function of a single effective parameter. This concludes the proof of Eq. \eqref{eq:ideal_Ec_twoqb}.


\subsection[{Energetics of the g-JQB}]{Energetics of the $g$-JQB}
\label{app:null_Ec_glob}

In the $g$-JQB, the resource for extracting work comprises both the subsystems and the correlations. The state from which the global ergotropy $\mathcal{E}^{(g)}$ is extracted is the thermal state $\varrho_{\rm II}=\varrho_{\rm I} = \tau_\beta$ with Hamiltonian $H=H_0 + \gamma_0 H_{\rm int}$ \eqref{eq:Ham_gamma}. The purpose of this protocol is to extract $\mathcal{E}^{(g)}$ with respect to the bare Hamiltonian $H_0= \sigma_A^z +  \sigma_B^z$. Indeed, $\tau_\beta$ is passive with respect to $H = H_0+\gamma_0 H_{\rm int}$ but can be active with respect to $H_0$. Therefore, disconnecting the qubits is still a necessary condition for the working cycle based on this protocol. When ergotropy is globally extracted from a system, 
the passive state 
is not unique if there are degeneracies in the energy spectrum  \cite{Allahverdyan_epl}. 

In the present case, ergotropy is extracted from $\tau_\beta$ with respect to $H_0=\rm{diag}(2,0,0,-2)$, diagonal in the two-qubit computational basis, $\{\vert 00 \rangle,\vert 01 \rangle,\vert 10 \rangle,\vert 11 \rangle\}$. The two eigenstates associated with the degenerate zero-energy level are $\vert 01\rangle$ and $\vert 10 \rangle$. The presence of a degenerate eigenvalue makes the eigendecomposition of $H_0$ not unique (arbitrariness in choosing orthonormal states in the degenerate subspace), which, in turn, results in ambiguity in the definition of the unitary $\mathcal{U}_g$ in Eq. \eqref{eq:U_ergextr} by means of which one would like to extract the global ergotropy $\mathcal{E}^{(g)}$. Below, we show that this ambiguity does not affect the global ergotropy $\mathcal{E}^{(g)}$ and that, in the present model ($g$-JQB), $E_c^{(g)} \equiv 0$.

The aforementioned arbitrariness can be encoded in a unitary operator $u$, such that $[H_0, u]=0$, which effectively acts only on the degenerate subspace of $H_0$ spanned by $\{\vert 01 \rangle,\vert 10 \rangle\}$, and so that $u \vert 00 \rangle = \vert 00 \rangle$ and $u \vert 11 \rangle = \vert 11 \rangle$. In the computational basis, this operator can be represented by
\begin{equation}
u = 
\left(\begin{array}{cccc}
1 & 0 & 0 & 0 \\
0 & u_{01,01} & u_{01,10} & 0 \\
0 & u_{10,01} & u_{10,10} & 0 \\
0 & 0 & 0 & 1
\end{array}\right),
\end{equation}
with $uu^\dagger = u^\dagger u = I$. Accordingly, we introduce the unitary (see Eq. \eqref{eq:U_ergextr})
\begin{equation}
\mathcal{U}_{g;u} = u \sum_{k=1}^{4} \vert \epsilon_{k}^{(0)\uparrow} \rangle\langle \lambda_{k}^\downarrow \vert \equiv u \mathcal{U}_g,
\label{eq:Uu}
\end{equation}
where the eigenstates of $H_0$, $\{ \vert \epsilon_{k}^{(0)\uparrow} \rangle \}_k$ with associated eigenvalues  $\{\epsilon_{k}^{(0)\uparrow}\}= \{-2,0,0,2\}$, are the computational basis states sorted in reverse order (from the lowest energy state $\vert \epsilon_{1}^{(0)\uparrow} \rangle = \vert 11 \rangle$ to the highest one $\vert \epsilon_{4}^{(0)\uparrow} \rangle = \vert 00 \rangle$),  $\{ \vert \lambda_k^\downarrow \rangle \}_k$ the eigenstates of $\tau_\beta$. By construction, the unitary \eqref{eq:Uu} is expected to extract the global ergotropy, but its definition depends on $u$. Starting from the orthonormal eigenstates $\vert 01 \rangle$ and $\vert 10 \rangle$, the effective purpose of $u$ is to span orthonormal pairs of states---a unitary transformation preserves norm and inner product---from the degenerate subspace associated with $\epsilon_{2}^{(0)\uparrow}=\epsilon_{3}^{(0)\uparrow}=0$, without affecting $\vert 00 \rangle$ and $\vert 11 \rangle$. 

The arbitrariness of $\mathcal{U}_{g;u}$ \eqref{eq:Uu} results in an ambiguity in the passive state $\pi_{g;u} = u \pi_g u^\dagger$, where $\pi_g=\rm{diag}(\lambda_4^\downarrow,\lambda_3^\downarrow,\lambda_2^\downarrow,\lambda_1^\downarrow)$ is the diagonal state obtained by using the identity in place of $u$. In the computational basis, the passive state is of the form
\begin{equation}
\pi_u = 
\left(\begin{array}{cccc}
{\lambda}_{4}^\downarrow & 0 & 0 & 0 \\
0 & (\pi_u)_{01,01} & (\pi_u)_{01,10} & 0 \\
0 & (\pi_u)_{10,01} & (\pi_u)_{10,10} & 0 \\
0 & 0 & 0 & {\lambda}_{1}^\downarrow
\end{array}\right).
\end{equation}
The global ergotropy $\mathcal{E}^{(g)}(\tau_\beta) \equiv \Tr[H_0 \tau_\beta]-\Tr[H_0 \pi_g]$ turns out to be independent of $u$, because $\Tr[H_0\pi_u]=\Tr[H_0\pi_g]$ since $[H_0,u]=0$. In particular, it reads
\begin{eqnarray}
    \mathcal{E}^{(g)} &= \Tr[(\sigma_A^z+\sigma_B^z) \tau_\beta] - \sum_{k=1}^4 \lambda_k^\downarrow \epsilon_{k}^{(0)\uparrow}\nonumber\\
    &= 2 \langle \sigma_A^z \rangle -2(\lambda_4^\downarrow-\lambda_1^\downarrow),
\end{eqnarray}
where $\lambda_1^\downarrow = e^{-\beta \epsilon_1^\uparrow}/Z$ and $\lambda_4^\downarrow=e^{-\beta \epsilon_4^\uparrow}/Z$. In the derivation, we have used $\mathcal{U}_g$ in Eq. \eqref{eq:Uu}, $H_0 = \sum_{k=1}^4 \epsilon_{k}^{(0)\uparrow} \vert \epsilon_{k}^{(0)\uparrow} \rangle\langle \epsilon_{k}^{(0)\uparrow} \vert$ with $\{\epsilon_{k}^{(0)\uparrow}\}= \{-2,0,0,2\}$, and  $\tau_\beta = \sum_{k=1}^4 \lambda_k^\downarrow \vert \lambda_k^\downarrow \rangle\langle  \lambda_k^\downarrow \vert$, whose eigenvalues $\lambda_k^\downarrow = e^{-\beta \epsilon_k^\uparrow}/Z$ involve the partition function $Z = \Tr[e^{-\beta H}]$ of the Hamiltonian $H=H_0+\gamma_0 H_{\rm int}$, whose eigenvalues $\{\epsilon_k^\uparrow\}_k$ are given in \ref{app:Hspectrum}. This concludes the proof of $\mathcal{E}^{(g)}$ in Eq. \eqref{eq:ideal_Ergo_global}.

Now, we prove  that $E_c^{(g)} = \gamma_0 \Tr[H_{\rm int} \pi_u]\equiv 0$, see Eq. \eqref{eq:Ec}. In the computational basis, $H_{\rm int} = - (\sigma_A^{x} + \sigma_B^{x} + \sigma_A^x\sigma_B^x - \sigma_A^y\sigma_B^y)$ is of the form
\begin{equation}
H_{\rm int} =
\left(\begin{array}{cccc}
0			& {h}_{00,01} & {h}_{00,10} & {h}_{00,11} \\
{h}_{01,00} &	0				& 0					 & {h}_{01,11} \\
{h}_{10,00} &	0				& 0					 & {h}_{10,11} \\
{h}_{11,00} & {h}_{11,01} & {h}_{11,10} & 0\\
\end{array}\right).
\end{equation}
By direct inspection, regardless of the detailed definition of the matrix elements, we observe that the diagonal elements of the matrix $H_{\rm int} \pi_u$ are identically null, which trivially proves that $E_c^{(g)} =\Tr[H_{\rm int} \pi_u] = 0$.

In conclusion, considering the above results, for the $g$-JQB we simply use $\mathcal{U}_g$, since we can neglect $u$ in Eq. \eqref{eq:Uu} as well as the arbitrary phases of Eq. \eqref{eq:U_ergextr}, the ergotropy being extracted globally.


\section{Energy spectrum of the two-qubit Hamiltonian \eqref{eq:Ham_gamma}}
\label{app:Hspectrum}
In the computational basis, the matrix representation of the Hamiltonian \eqref{eq:Ham_gamma} is
\begin{equation}
H  = 
\left(\begin{array}{cccc}
2		    & -\gamma_0	& -\gamma_0	& -2\gamma_0\\
-\gamma_0	& 0			& 0			& -\gamma_0\\
-\gamma_0	& 0			& 0			& -\gamma_0\\
-2\gamma_0  & -\gamma_0	& -\gamma_0	& - 2
\end{array}\right),
\end{equation}
whose characteristic polynomial reads
\begin{equation}
p_H(\epsilon)=\det( H - \epsilon I ) = \epsilon[\epsilon^3-4(1+2\gamma_0^2)\epsilon+8\gamma_0^3].
\end{equation}
Therefore, the (unsorted) energy levels are
\begin{equation}
\{ \epsilon_k \}_{k=1,\ldots,4} = \left\{0, \frac{4\sqrt{1+2\gamma_0^2}}{\sqrt{3}}\cos\left[\frac{\phi+2\pi n}{3}\right] \right\}
\label{eq:specH}
\end{equation}
with $n=0,1,2$ and 
\begin{equation}
\phi = \arccos\left[-\frac{3\sqrt{3}\gamma_0^3}{2(1+2\gamma_0^2)^{3/2}}\right].
\end{equation}
The three real roots of the depressed cubic equation, $\epsilon^3-4(1+2\gamma_0^2)\epsilon+8\gamma_0^3=0$, are obtained via a trigonometric approach \cite{Zucker_2008}. Note that the lowest, $\epsilon_1^\uparrow$, and the highest, $\epsilon_4^\uparrow$, energy levels are respectively associated with $n=1$ and $n=0$, see Fig. \ref{fig:spectrumH}. Notice that the determination of the analytical expression of the eigenstates of $H$ is hindered by the cumbersome expression of the eigenvalues.

\begin{figure}[!tb]
\centering
    \includegraphics[width=0.7\columnwidth]{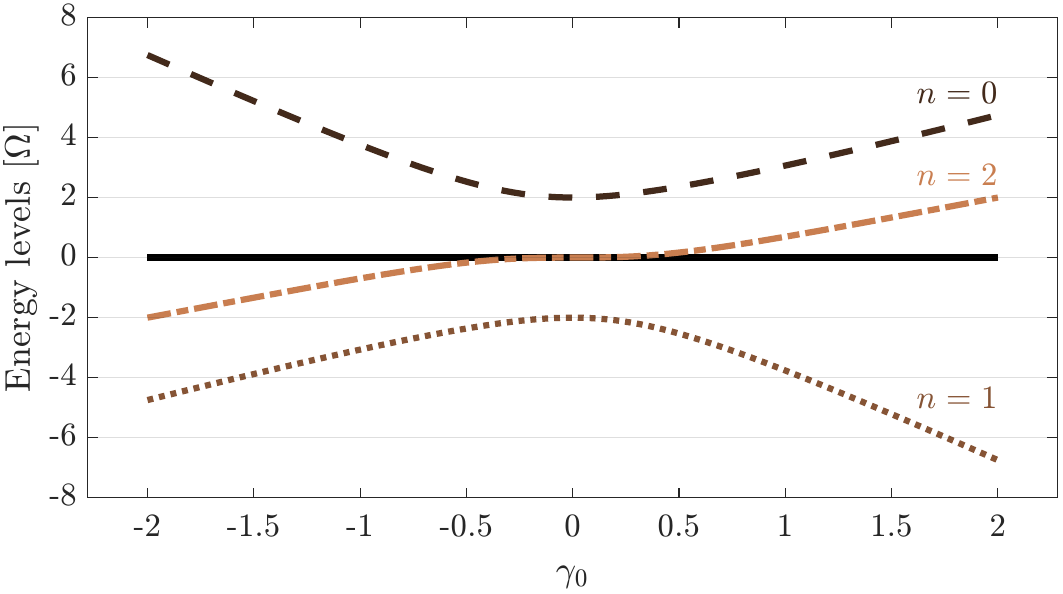}
    \caption{Spectrum \eqref{eq:specH} of the two-qubit Hamiltonian \eqref{eq:Ham_gamma}.
    \label{fig:spectrumH}}
\end{figure}


\section{Entanglement and ergotropic gap}
\label{app:entg_erggap}
In the main text, we have shown that the possibility of enhancing the performance of the quantum battery is related to the value of the relative phase in the local unitary \eqref{eq:Uerg_qiskit}---revealing the role of coherence---together with the fact that strokes \textsc{(i$\rightarrow\ldots\rightarrow$iv)} preserve quantum correlations
between the qubits 
{when ergotropy is extracted by local unitaries ($p=s,l$).} In this appendix, we analyze the role of entanglement. After introducing and reviewing the quantities of interest, we review how entanglement depends on the parameters of the model, $(\gamma_0, T)$, and, we relate it with the ergotropy locally extracted from a single qubit and with the ergotropic gap \cite{Mukherjee_pre2016}, which is the difference between the global and the local ergotropy.


\subsection{Definitions}
For a two-qubit state $\varrho$, the entanglement of formation can be measured as \cite{Wootters_prl}
\begin{equation}
E_F(\varrho) = h\left(\frac{1+\sqrt{1-C^2(\varrho)}}{2}\right),
\label{eq:entg_conc}
\end{equation}
where $h(x)= - x \log_2 x -(1-x)\log_2(1-x)$ is the binary Shannon entropy associated with the probability distribution $\{x,1-x\}$. The quantity $C(\varrho)=\max\{0,\mu_1^\downarrow-\mu_2^\downarrow-\mu_3^\downarrow-\mu_3^\downarrow\}$ is the \textit{concurrence} of the state $\varrho$, where $\{\mu_k^\downarrow\}$ are the square roots of the eigenvalues, in descending order, of the Hermitian matrix $R = \sqrt{\varrho} \tilde{\varrho} \sqrt{\varrho}$ (or equivalently of the non-Hermitian matrix $R'=\varrho \tilde{\varrho}$).
Here, $\tilde{\varrho}=(\sigma_y \otimes \sigma_y)\varrho^\ast (\sigma_y \otimes \sigma_y)$ denotes the spin-flipped state and the complex conjugation is taken in the computational basis of the two-qubits. Each $\mu_k^\downarrow$ is a non-negative real number. The measure $E_F(\varrho)$ is a monotonically increasing function of $C(\varrho)$ and ranges from 0 (no entanglement) to 1 (maximum entanglement) as $C(\varrho)$ goes from 0 to 1.

In a multipartite scenario, the presence of quantum correlations (including but not limited to entanglement) is always associated to a nonvanishing ergotropic gap \cite{Mukherjee_pre2016,Alimuddin_pra2019,Perarnau-Llobet_prx2015}, defined as the difference between the \textit{global} and the \textit{local} ergotropy 
\begin{equation}
\Delta_{\rm EG} \equiv \mathcal{E}^{(g)}-\mathcal{E}^{(l)}.    
\end{equation}
The converse, however, does not necessarily hold, meaning that a nonzero ergotropic gap does not imply the presence of quantum correlations: There exist indeed classically correlated states for which $\Delta_{\rm EG} \neq 0$ \cite{Mukherjee_pre2016}. In detail, letting $\varrho$ be the state of a multipartite system with Hamiltonian $H = \sum_{k}H_k$, where each local Hamiltonian $H_k$ acts on the $k$-th subsystem with reduced state $\rho_k$, the \textit{global} and the \textit{local} ergotropy that can be extracted from $\varrho$ with respect to $H$ are, respectively,
\begin{eqnarray}
\mathcal{E}^{(g)} &=\Tr[H \varrho]- \min_{\mathcal{U}_g} \Tr[H \mathcal{U}_g \varrho\, \mathcal{U}_g^\dagger] ,\\
\mathcal{E}^{(l)} &=\Tr[H \varrho]- \min_{\mathcal{U}_l} \Tr[H \mathcal{U}_l \varrho\, \mathcal{U}_l^\dagger] ,
\end{eqnarray}
where $\mathcal{U}_l = \bigotimes_{k}U_k$ is the tensor product of local unitaries each of which acts on a single subsystem. Then, the ergotropic gap reads $\Delta_{\rm EG}= \sum_{k} \Tr[H_k \pi_k] - \Tr[H \pi_g]$, where $\pi_g = \mathcal{U}_g \varrho\, \mathcal{U}_g^\dagger$ is the passive state of $\varrho$ with respect to $H$ and $\pi_k = U_k \rho_k U_k^\dagger$ is the passive state of $\rho_k$ with respect to $H_k$. The ergotropic gap is nonnegative, $\Delta_{\rm EG}\geq 0$, because local unitaries can extract work from subsystems only, while the global unitary, in addition, can extract it from correlations \cite{Mukherjee_pre2016}.

In this work, we consider a two-qubit system with Hamiltonian \eqref{eq:Ham_gamma}, given in an initial thermal state. After disconnecting the two qubits ($\gamma(t)=0$), the purpose is to extract the local ergotropy $\mathcal{E}^{(l)}=2\mathcal{E}$, see Eq. \eqref{eq:ideal_Erg},  
or the global ergotropy $\mathcal{E}^{(g)}$ \eqref{eq:ideal_Ergo_global} from the two-qubit thermal state with respect to the bare Hamiltonian $H_0 = H_A + H_B$, where the local Hamiltonians are $H_k=\sigma_k^z$, with $k=A,B$. In this context,
the ergotropic gap reads
\begin{equation}
\Delta_{\rm EG} = \frac{2}{Z}\left( e^{-\beta \epsilon_1^\uparrow} - e^{-\beta \epsilon_4^\uparrow} \right)-2r,
\label{eq:erggap_analytical}
\end{equation}
where $r = \sqrt{x^2+y^2+z^2}$ (see Eq. \eqref{eq:bloch_coords} and recall $y=0$ for real states), $Z = \Tr[e^{-\beta H}]$ is the partition function and the eigenvalues $\{\epsilon_{k}^\uparrow\}$ of $H  = H_0 + \gamma_0 H_{\rm int}$ are given in \ref{app:Hspectrum}. The lowest, $\epsilon_1^\uparrow$, and the highest, $\epsilon_4^\uparrow$, energy eigenvalues are respectively associated with $n=1$ and $n=0$ in Eq. \eqref{eq:specH}.

\begin{figure}[!tb]
	\centering
	\includegraphics[width=0.7\columnwidth]{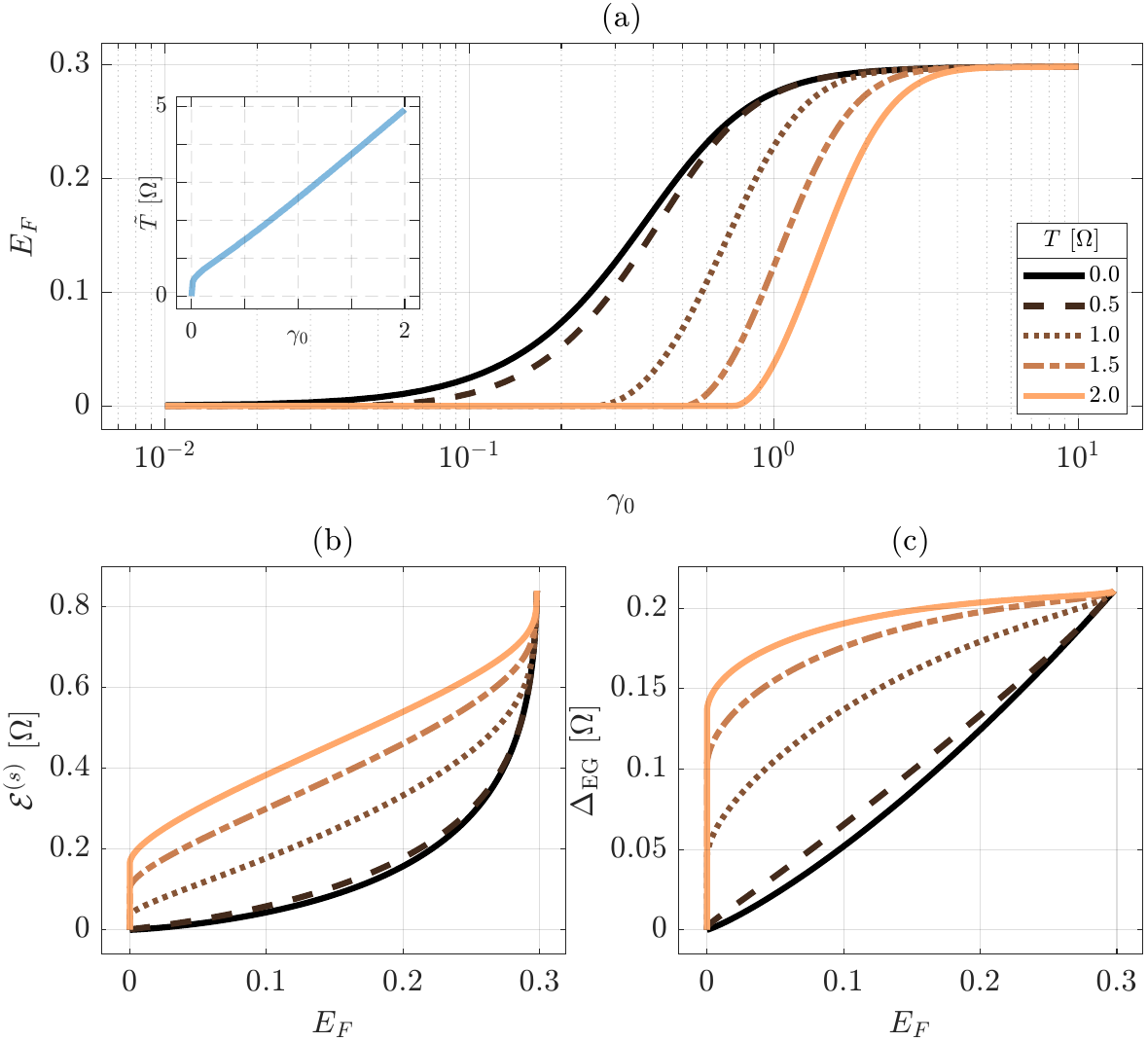}
	\caption{(a) Entanglement of formation $E_F$ \eqref{eq:entg_conc} of the two-qubit state 
    {before work extraction} as a function of $\gamma_0$. The inset shows, for each value of $0 \leq \gamma_0 \leq 2$, the threshold temperature $\tilde{T}$ above which entanglement vanishes, $E_F=0$ for $T>\tilde{T}$. (b) Ergotropy $\mathcal{E}^{(s)}$ extracted from a single qubit in the $s$-JQB and (c) ergotropy gap $\Delta_{\rm EG}=\mathcal{E}^{(g)}-\mathcal{E}^{(l)}$ \eqref{eq:erggap_analytical} as a function of $E_F$. Results are reported for different temperatures.
    }
	\label{fig:entg_1qbbat}
\end{figure}


\subsection{Discussion}
{The entanglement of the state before work extraction is the same as that of the initial thermal state, $E_F(\tau_\beta)$, regardless of the protocol for work extraction, because the thermal state is preserved through the instantaneous disconnection stroke \textsc{(i$\rightarrow$ii)}. Our purpose is to investigate how such initial entanglement depends on the parameters of the system, $\gamma_0$ and $T$, and then to relate it with the ergotropy locally extracted from the single qubit and the ergotropic gap.} Here, we focus on $\gamma_0 > 0$, the regime in which the quantum battery performs better (see Fig. \ref{fig:idperf_LG_qbbat}).

In the regime of interest, entanglement is a non-decreasing function of $\gamma_0$, which is the coupling strength of the interaction [Fig. \ref{fig:entg_1qbbat}(a)]. At low temperature, small values of $\gamma_0$ suffice to establish entanglement, while at higher temperature, larger values are required and entanglement reaches the asymptotic value $E_F \approx 0.3$ for $\gamma_0 \gg 1$. The inset shows that for each value of $\gamma_0$, there exists a critical temperature $\tilde{T}$ above which the entanglement of formation vanishes. Except for $\gamma_0 \ll 1$, $\tilde{T}$ linearly depends on $\gamma_0$.

Focusing on the extractable work, finite ergotropy is locally extracted from the single qubit in the presence of entanglement at low temperature, as well as at high temperature in the absence of entanglement [Fig.~\ref{fig:entg_1qbbat}(b)].

The study of the ergotropic gap is consistent with the expected behavior: 
At $T=0$ the thermal state---the ground state of $H$---shows finite entanglement (purely quantum correlation) and non-vanishing $\Delta_{\rm EG}$; for $T>0$,  
we can observe non-vanishing $\Delta_{\rm EG}$ in the absence of entanglement [Fig. \ref{fig:entg_1qbbat}(b)]. Finally, recalling that entanglement is a non-decreasing function of $\gamma_0$, we observe that larger values of entanglement are accompanied by larger values of the ergotropic gap.

\section*{References}
\bibliography{cycssqb_biblio}

\end{document}